\documentclass[acmsmall]{acmart}

\AtBeginDocument{%
  }

\setcopyright{acmlicensed}
\copyrightyear{2018}
\acmYear{2018}
\acmDOI{XXXXXXX.XXXXXXX}

\acmJournal{JACM}
\acmVolume{37}
\acmNumber{4}
\acmArticle{111}
\acmMonth{8}

\usepackage[english]{babel}
\usepackage[T1]{fontenc}
\usepackage[utf8]{inputenc}

\usepackage{multirow}
\usepackage{amsmath}
\usepackage{caption}
\usepackage{xspace}
\usepackage[skip=0cm,list=true,labelfont=it]{subcaption}
\usepackage{gensymb}
\usepackage{enumitem}
\usepackage{algorithm}
\usepackage{algpseudocode}
\usepackage{tikz}

\newcommand{\sysname}{$\sf\small{EyeNexus}$\xspace}

\newcommand{\circlednumber}[1]{\tikz[baseline=(char.base)]{
    \node[shape=circle, draw, inner sep=2pt] (char) {#1};}}

\usepackage{mdframed}
\definecolor{lavendergray}{rgb}{0.77, 0.76, 0.82}
\definecolor{lightgray}{rgb}{0.88, 0.88, 0.88}
\newmdenv[
  backgroundcolor=lightgray,
  linecolor=lightgray,
  linewidth=1pt,
  leftmargin=4pt,
  rightmargin=2pt,
  innertopmargin=4pt,
  innerbottommargin=2pt,
  innerleftmargin=2pt,
  innerrightmargin=2pt
]{takeaway}

\hypersetup{
    colorlinks=true,
    linkcolor=blue,
    filecolor=magenta,
    urlcolor=cyan,
    citecolor=blue
}

\let\OldS\S
\renewcommand{\S}{\OldS\xspace}

\usepackage{titlesec}
\titlespacing{\section}{30pt}{\parskip}{-\parskip}
\titlespacing{\subsection}{10pt}{\parskip}{-\parskip}
\titlespacing{\subsubsection}{10pt}{\parskip}{-\parskip}

\begin{document}

% \title[Gaze-Guided VR Game Streaming]{Gaze-Guided Streaming: Harnessing the Human Visual System for Seamless VR Cloud Gaming Experiences}

\title[Gaze-Guided VR Game Streaming]{EyeNexus: Adaptive Gaze-Driven Quality and Bitrate Streaming for Seamless VR Cloud Gaming Experiences}
\settopmatter{authorsperrow=5} % optional: control authors per row

\author{Ze Wu}
\authornote{Co-first author}
\affiliation{
  \institution{Hong Kong University of Science and Technology}
  \city{Hong Kong}
  \country{China}
}
\email{zwubn@connect.ust.hk}

\author{Ahmad Alhilal}
\authornotemark[1] % shares the same authornote (co-first)
\affiliation{
  \institution{Aalto University}
  \country{Finland}
}
\email{ahmad.alhilal@aalto.fi}

\author{Yuk Hang Tsui}
\authornotemark[1]
\affiliation{
  \institution{Hong Kong University of Science and Technology}
  \city{Hong Kong}
  \country{China}
}
\email{yhtsui@connect.ust.hk}

\author{Matti Siekkinen}
\affiliation{
  \institution{Aalto University}
  \country{Finland}
}
\email{matti.siekkinen@aalto.fi}

\author{Pan Hui}
\authornote{Pan Hui is also affiliated with the Hong Kong University of Science and Technology, Hong Kong SAR, and the University of Helsinki, Finland.}
\affiliation{
  \institution{Hong Kong University of Science and Technology (Guangzhou)}
  \city{Guangzhou}
  \country{China}
}
\email{panhui@ust.hk}
\renewcommand{\shortauthors}{Wu et al.}
\begin{abstract}
Virtual Reality (VR) cloud gaming systems render the 3D graphics on cloud servers for playing graphically demanding games on VR headsets. Delivering high-resolution game scenes is challenging due to variation in network performance.
By leveraging the non-uniform human vision perception, foveated rendering and encoding have proven effective for optimized streaming in constrained networks.
SoTA foveation methods either do not incorporate real-time gaze data or are unable to handle variations in network conditions, resulting in a suboptimal user experience.
We introduce \sysname, a pioneering system that combines real-time gaze-driven spatial compression (FSC) with gaze-driven video encoding (FVE), transforming the gaze point for precise alignment and foveation. We propose a novel foveation model that dynamically adjusts the foveation region based on real-time bandwidth and gaze data. The model simplifies network-aware quality assignment in FVE, ensuring smooth and imperceptible quality gradients.
We evaluate \sysname using objective and subjective measures with different network conditions and games. \sysname reduces latency by up to 70.9\% and improves perceptual visual quality by up to 24.6\%. 
Our IRB-approved user study shows that \sysname achieves the highest playability and visual quality, with improvements of up to 48\%, while eliminating motion sickness.

\end{abstract}

% \ccsdesc[100]{Computer systems organization~Real-time system architecture}

\begin{teaserfigure}
  \includegraphics[width=\textwidth]{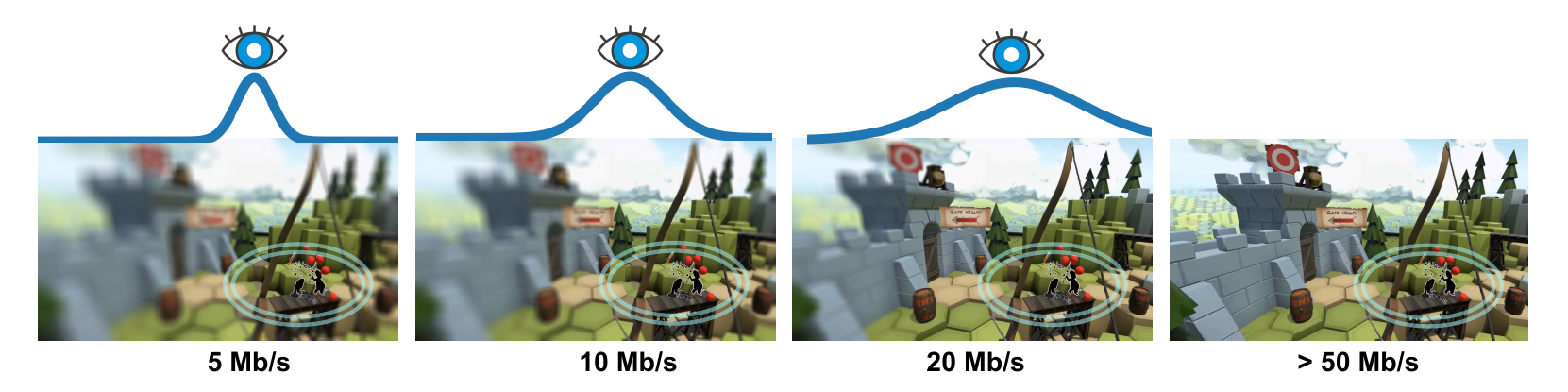}
  \vspace{-.6cm}
  \caption{\sysname's quality allocation, driven by gaze and available bandwidth.}
  \label{fig:teaser}
\end{teaserfigure}


\maketitle

\section{Introduction}
The Virtual Reality (VR) gaming market is poised for substantial growth, driven by significant investments from companies such as Meta, Google, HTC, Sony, and Samsung. It is projected to reach USD 189.17 billion by 2032~\cite{VRgamingMarket2024}. Mobile (wireless) VR headsets enhance the gaming experience by streaming high-quality content without physical connections, allowing gamers to move freely and increasing their sense of presence. Cloud VR gaming offloads rendering tasks to powerful cloud servers, eliminating the need for expensive high-end gaming PCs. These servers process user motion data and render game scenes, streaming them as 2D video frames over the internet. Real-time streaming of cloud-rendered frames requires a high-speed internet connection and adequate bandwidth. However, wireless and mobile networks often face fluctuations in latency, throughput, and packet loss. Moreover, ensuring a good quality of experience (QoE) in cloud VR is challenging, as users expect high visual quality, low response times, and minimal cybersickness.

The state-of-the-art addresses network challenges in cloud gaming in two ways: 1) adaptive bitrate (ABR), which adjusts video bitrate based on network performance ( bandwidth, latency, and packet loss), and 2) foveation, which leverages the non-uniform acuity of the human visual system to reduce computation load and bandwidth demand. 
% ABR Streaming 
In terms of ABR streaming, WebRTC~\cite{webrtc} is an industry-standard developed by Google that uses Google Congestion Control (GCC)~\cite{GCCwebrtc} for bitrate selection. GCC favors real-time transmission to video quality. Nebula~\cite{alhilal2022nebula} adapts the video rate to cope with bandwidth variations and applies frame-level redundancy to avoid visual distortions. Air Light VR (ALVR)~\cite{alvr} is an open-source VR gaming system that streams VR games from high-end gaming PCs to VR headsets via Wi-Fi. ALVR adapts the target bitrate according to frame interval and bitrate. 
% HVS Streaming 
In terms of foveation, related works harness the human eye's uneven perception of visual quality~\cite{humanvisualsystem} through foveated rendering (FR) and foveated video encoding (FVE). FR reduces computational load by lowering peripheral resolution, while FVE reduces bitrate through aggressive compression in the periphery~\cite {alhilal2024FovOptix, Illahi2020CGFVE}. Specifically, cloud gaming with Foveated video encoding (CGFVE) leverages the non-uniformity in HVS to reduce the bandwidth requirement~\cite{Illahi2020CGFVE}. Only FovOptix~\cite{alhilal2024FovOptix} integrates foveation into 
rendering and video encoding. 
FovOptix assigns variable quality across three fixed FoV regions and adapts quality based on network performance.
%Research Gap
However, WebRTC and Nebula use uniform encoding and rendering for video streaming and mobile cloud gaming, resulting in suboptimal QoE. While ALVR and CGFVE partially integrate foveation, they lack adaptability to network fluctuations, leading to poor performance in mobile conditions. FovOptix is designed for a fixed first-person perspective and does not adapt to users' dynamic gaze shifts. Additionally, FovOptix applies uniform quality within each of the three fixed regions, resulting in noticeable quality gradients.

%****** Contribution **********************
In this paper, we propose \sysname, a pioneering system that combines gaze-driven spatial compression (FSC) with adaptive gaze-driven video streaming (FVE) in VR cloud gaming. FSC compresses peripheral pixels while maintaining full resolution in the foveal area, reducing encoding and decoding latency and optimizing bandwidth. FVE assigns the quantization (quality) non-uniformly across the game frame, guided by users' real-time gaze. Unlike the state of the art, we associate the quantization settings with the available bandwidth to optimize the quality of experience, as illustrated in~\autoref{fig:teaser}. Specifically, we simplify network-aware quality assignment by 
modeling a foveation controller following Gaussian distribution(bell curve) that 
ensures smooth quality gradients. Consequently, \sysname accounts for the non-uniform acuity of HVS and network variations while optimizing the visual experience. We develop  \sysname on ALVR opensource to interoperate with SteamVR~\cite{steamVR} and ensure game and platform independence. This allows players to install any VR game and supports various VR headsets. Our contribution is summarized as follows:
\begin{itemize}[leftmargin=0em,topsep=.5em]
    \item \textbf{Introduce} \sysname, the pioneering VR game streaming system that combines gaze-contingent spatial compression (FSC) with gaze-contingent video encoding (FVE).
    \item \textbf{Implement} a dynamic FSC utilizing real-time gaze tracking to lower pixel density in the peripheral region, for faster video coding and optimized bandwidth usage. 
    \item \textbf{Design} a foveation model for FVE to associate quality assignment with network performance, following Gaussian distribution for seamless streaming and enhanced perceptual visual quality.
    \item \textbf{Integrate} SoTA benchmarks (research and industry standards) into an open-source mobile VR gaming platform.
    \item \textbf{Evaluate} \sysname against the benchmarks using objective and subjective measures. \sysname presents the lowest latency and enhanced perceptual image quality, improving the playability, perceived visual quality, and susceptibility to motion sickness.  
\end{itemize}

% The remainder of the paper is structured as follows. In \S \ref{sec:related}, we summarize the related work. We discuss the methodology of \sysname in section \ref{sec:sysdesign}, and its detailed implementation in section \ref{sec:impl}. After presenting the experiment setup in \ref{sec:setup}, we present the results in section \ref{sec:eval}. Finally, we conclude our work in section \ref{sec:conclusion}.

\section{Background and Related Work} 
\label{sec:related}

The state-of-the-art addresses network fluctuations in cloud gaming in two ways: 1) Adaptive bitrate (ABR) streaming that adjusts the sending bitrate according to network performance and 2) Reducing bandwidth demand by leveraging the non-uniform perception in human vision.

%%%% Subsection
\subsection{Adaptive Bitrate in Cloud Gaming}
Google Congestion Control (GCC)~\cite{GCCwebrtc} is an ABR approach that selects the bitrate based on observations of the network throughput, delay gradients, and fraction packet loss. WebRTC~\cite{webrtc}, Google media streaming standard, utilizes GCC for real-time streaming. However, a drawback of GCC is its preference for real-time transmission over video quality \cite{gccAnalysis}. Air Light VR (ALVR)~\cite{alvr} is an open-source system that streams VR games from high-end gaming PC to VR headsets via Wi-Fi. In ALVR, users have the option to choose between constant bitrate and ABR mode. In ABR mode, ALVR tracks the frame interval time and the sending bitrate for each frame, storing these values in a sliding window, and adjusts the target bitrate based on their average values. Nebula~\cite{alhilal2022nebula} is an ABR approach in mobile cloud gaming. Nebula adapts the bitrate to cope with bandwidth variations with frame-level redundancy to avoid visual distortions. 

While GCC and Nebula consider network conditions, they are tailored for traditional video streaming and cloud gaming. VR cloud gaming faces greater challenges due to display proximity, ultra-high resolutions, and low latency needs. ALVR requires a stable network to provide a satisfactory performance. \sysname introduces dynamic and adaptive foveation with rate control to optimize interaction and immersion under fluctuations of network conditions.

\begin{figure}[t]
    \centering
    \begin{subfigure}[b]{\textwidth}
        \includegraphics[width=0.95\linewidth]{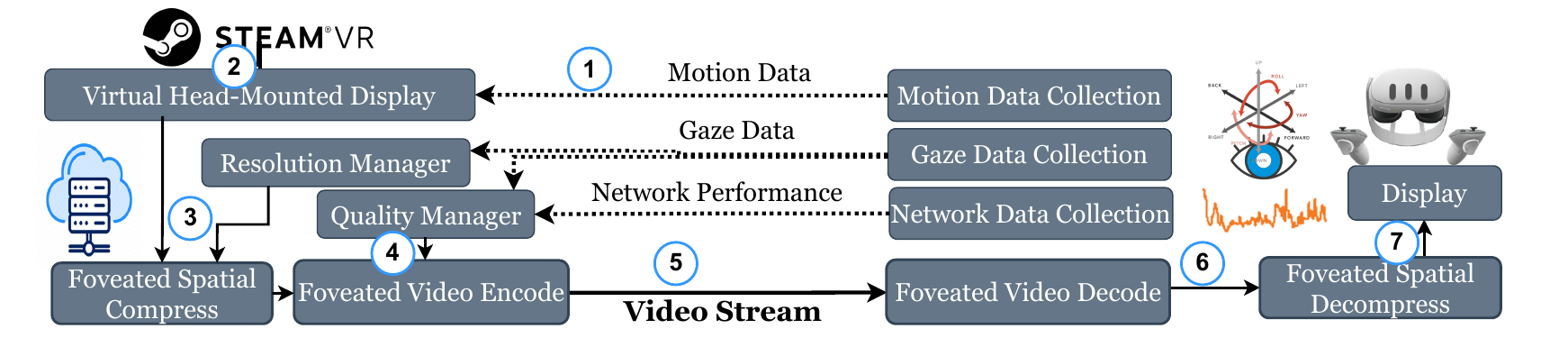}
        \caption{End-to-end architecture }
    \label{fig:architecture}
    \end{subfigure}
    \begin{subfigure}[b]{\textwidth}
        \includegraphics[width=0.95\linewidth]{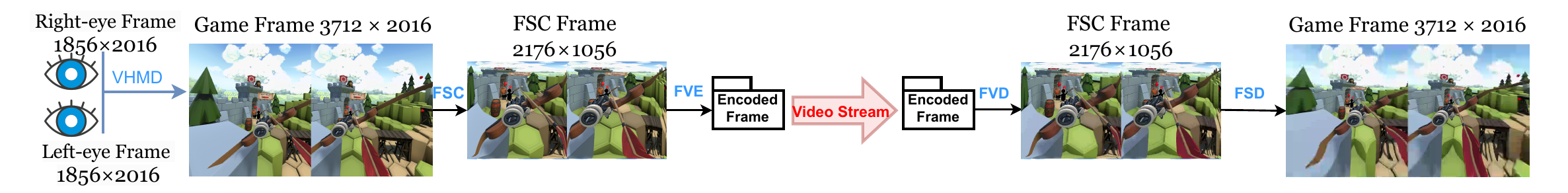}
        \caption{Frame transformation over the pipeline }
        \label{fig:frames_over_arch}
    \end{subfigure}
    \caption{\sysname's end-to-end architecture and corresponding frame transformation}
    \label{fig:sysarch}
\end{figure}

%%%% Subsection
\subsection{Foveated Cloud Gaming and VR}
Foveated rendering has been studied with VR (e.g., ~\cite{patney16foveation, walton21metamers, rolff23vrsnerf}). However, the goal has been to reduce the computational load of rendering by reducing the workload in the shading phase of the rendering pipeline, for instance, whereas we aim to reduce bitrate and latency when streaming foveated graphics. In this sense, those solutions can be seen as complementary to our work. 

In this work, we focus on foveated spatial compression (FSC) to reduce the number of pixels through 2D warping done through post-processing. The goal is to reduce the workload of video encoding and transmission when streaming those graphics, which makes it different from the traditional foveated rendering approach mentioned above. This type of foveation has been studied earlier in~\cite{illahi21mmsys} where it was compared to foveated video encoding (FVE) in which quantization of macroblocks is varied within a frame in a gaze-contingent manner and in~\cite{kamarainen23metasys} it is further combined with superresolution. Both spatial compression and FVE reduce bitrate, but the former also has the advantage of reducing latency. %\ze{To be deleted: In this work, we combine these approaches in an adaptive way.}

Concerning other work in foveated cloud gaming and VR, Illahi et al.~\cite{Illahi2020CGFVE} propose CGFVE, which explores the parameterization of foveated quantization but did not consider rate control to handle variable bandwidth scenarios and adapt the foveation. FovOptix~\cite{alhilal2024FovOptix} is a streaming method designed for VR cloud gaming that aligns with a fixed human vision perspective. It combines foveated rendering (FR) with an adaptive FVE to deliver a video stream at a lower yet adaptive bitrate. However, it is designed for first-person/perspective gaming by always assuming center fixation, thereby neglecting gaze dynamics altogether, whereas EyeNexus adjusts foveation based on real-time gaze tracking. In ~\cite{fang23nossdav}, the authors show that dynamic foveation in cloud VR improves user experience. However, they did not consider bandwidth-varying scenarios at all and, consequently, the foveation approach is non-adaptive without rate control, whereas EyeNexus introduces dynamic and adaptive foveation with rate control. 

%Illahi et al.~\cite{Illahi2020CGFVE} propose CGFVE, Foveated Video Encoding (FVE) for cloud gaming. CGFVE reduces the bandwidth requirement by taking advantage of the non-uniform acuity of the human visual system and by knowing where the user is looking. FovOptix~\cite{alhilal2024FovOptix} is a streaming method designed for VR cloud gaming that aligns with a fixed human vision perspective. FovOptix combines fixed FR with an adaptive fixed FVE to deliver video stream at a lower yet adaptive bitrate. FovOptix maintains perceived video quality in first-person/perspective view VR games without compromise. 

%\gap{FovOptix employs a fixed-foveated approach that does not account for the dynamic nature of eye gaze direction, where gaze shifts frequently from the FoV center in non-first-person perspective VR games. FovOptix assigns a single quality value to each of three fixed regions (center, near-periphery, periphery), resulting in noticeable quality gradients. Conversely, CGFVE treats PCs as client devices and performs well at resolutions under 2K. However, its performance drops significantly at 2K or higher in VR games due to the lack of foveated rendering, leading to increased video encoding delays.}
%\ze{Furthermore, it is important to note that CGFVE relies on TCP for video streaming and lacks rate control at the application layer. While their methodology in FVE significantly reduces bandwidth requirements, their solution does not ensure a high-quality experience in mobile network conditions.}

\section{System Design} 
\label{sec:sysdesign}

\sysname first applies foveated spatial compression (FSC) as a post-processing step to reduce pixel density and frame resolution, preserving resolution around the gaze point while lowering it in the periphery. Subsequently, \sysname assigns quality non-uniformly across the FSC frame in foveated video encoding (FVE), adapting to network variations. Unlike CGFVE, the quality assignment is time-varying and linked to the available bandwidth. In contrast to FovOptix, which employs fixed foveated rendering based on a first-person perspective, \sysname's FSC and FVE use real-time gaze points. In FSC, \sysname converts the gaze point from VR space to screen space coordinates, and in FVE, it transforms the gaze point from screen to FSC frame coordinates. This underscores the technical challenges of real-time gaze tracking not addressed by existing methods.

\subsection{Architecture Overview}
\label{sec:sysarch}

\autoref{fig:architecture} illustrates two instances, the client (VR headset application) and the high-end gaming server. The client encompasses data collection modules (network, eye gaze, and motion tracking), foveated video decoder (FVD), foveated spatial decompression (FSD), and display modules (for displaying the game frames). The server encompasses a virtual head-mounted device (VHMD), foveated spatial compression (FSC), foveated video encoding (FVE), and quality and resolution manager module. The VHMD module retrieves the physical pose and motion data from the client, and passes the data to SteamVR, and obtains a left-eye frame and a right-eye frame. VHMD then composites the frames into one game frame. FSC reduces frame resolution by compressing pixels in the peripheral region while preserving the resolution in the foveal region. FVE then encodes the FSC frame in a Gaussian distributional fashion, where the center around the eye gaze point receives the highest quality. %The encoded frame is transmitted to the client which receives the video packets, decodes them to recover the FR frame, applies reverse foveated rendering, and finally plays back the recovered frame. Simultaneously, the client monitors the player's physical pose and motion (i.e., hand controller input, head rotation, and viewing angle). The clients constantly transmit this data to the server to render and encode the next game frame.
\autoref{fig:frames_over_arch} illustrates the outcome of each module along the pipeline. 

\subsection{Gaze-driven Spatial Compression (FSC)}
\label{sec:gaze_rendering}
%this section, we demonstrate the computation of the eye gaze point's projection on the original game frame and the dynamic foveated rendering component employed in \sysname. The major contribution of foveated rendering is to reduce the frame size passing to the encoder, as the majority of encoders only support encoding frame size with resolution below $4096\times 4096$\cite{nvidia_video_codec_sdk}. To further take advantage of the foveated rendering process, we modified the foveated rendering to mapping based on gaze point, reducing the detriment on the QoE.

To enable real-time gaze-driven post-rendering, \sysname first maps VR gaze coordinates to screen space, and applies FSC to reduce the frame resolution sent to the encoder. This ensures faster encoding and supports higher resolutions (above 4K), as most hardware-accelerated codecs only handle up to 4K~\cite{nvidia_video_codec_sdk}.

\textbf{ \circlednumber{1} Gaze Mapping: VR to Screen Space Coordinates. } 
\begin{figure}[t!]
    \centering
    \begin{subfigure}[b]{.5\textwidth}
        \includegraphics[width=1.08\linewidth]{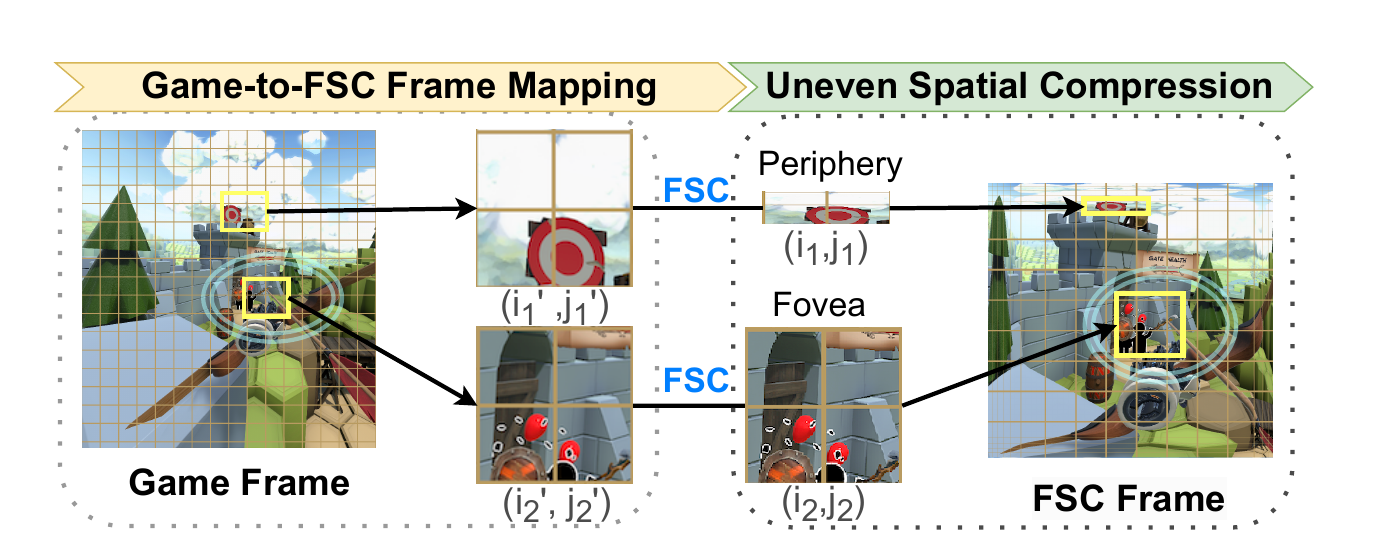}
        \caption{Process flow of foveated spatial compression (FSC) }
        \label{fig:FR_DFSC}
    \end{subfigure}
    \begin{subfigure}[b]{.47\textwidth}
        \includegraphics[width=\linewidth]{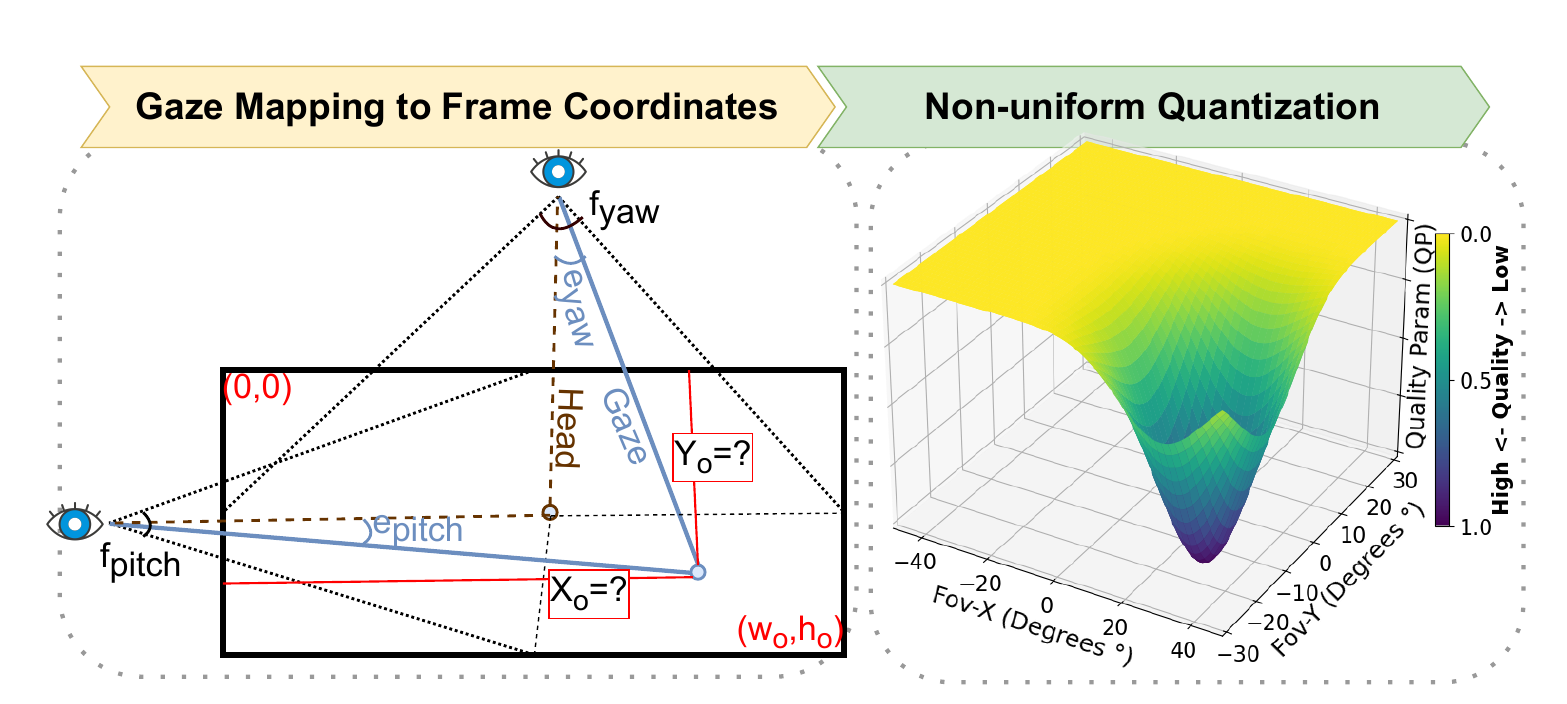}
        \caption{VR Space to 2D frame projection}
    \label{fig:Gaze_projection}
    \end{subfigure}

    \caption{ (a) Mapping and compression using AADT function, and (b)  VR 3D Space to 2D frame projection and illustration of non-uniform quantization.}
    \label{fig:FR_process}
\end{figure}

The VR headset collects the orientation of a gaze point in VR 3D space, represented as the vertical and horizontal angle of the gaze (\(f_{yaw},f_{pitch})\). Since FSC operates on 2D frames, we project the gaze angles into the screen space coordinates by computing the gaze point in the game frame, as demonstrated in~\autoref{fig:Gaze_projection}. The gaze point \((X_o,Y_o)\) are computed as follows: 
\begin{equation}
\begin{aligned}
&X_o = \frac{(\tan (\lvert f_{left} \rvert) - \tan (f_{yaw})) \cdot W_o}{\tan (\lvert f_{left} \rvert) + \tan (\lvert f_{right} \rvert)} \quad \text{,} \quad 
&Y_o = \frac{(\tan (\lvert f_{up} \rvert) - \tan (f_{pitch})) \cdot H_o}{\tan (\lvert f_{up} \rvert) + \tan (\lvert f_{down} \rvert)}
\end{aligned}
\end{equation}
Where \(f_{left}, f_{right}, f_{up}, f_{down}\) denote the field of view (FoV) angels in each direction, and \((W_o,H_o)\) denote the width and height of the game frame. 

\textbf{ \circlednumber{2} Dynamic Foveated Spatial Compression (FSC).} 
Existing FSC algorithms~\cite{foveatedspatialcompression} assume the frame's center as the foveal region, applying spatial compression to peripheral areas. However, frequent gaze shifts in non-first-person views, such as third-person shooting games, render this assumption inadequate. To address this limitation, we propose a dynamic FSC algorithm that adapts the fovea region to the user's real-time eye gaze location, optimizing perceptual visual quality and encoding efficiency.
We use Axis-aligned Distorted Transfer (AADT), utilized by Meta in the Oculus Link function~\cite{AADTOculus}, to implement the FSC algorithm. %\Arnold{The DFSC function is developed based on the \((W_o,H_o)\), the size of the region of interest (RoI) surrounding the gaze point ($X_{size}$, $Y_{size}$) and the horizontal and vertical compression ratios ($X_{comp}$, $Y_{comp}$). The dimension of the FSC Frame ($W_r, H_r$) is first computed, represented in \autoref{eq: FR dimension}}.
We first compute the dimension of the FSC frame ($W_r, H_r$) based on the game frame \((W_o,H_o)\), the size of the region of interest (RoI) surrounding the gaze point ($X_{size}$, $Y_{size}$) and the horizontal and vertical compression ratios ($X_{comp}$, $Y_{comp}$).
\begin{equation}
\begin{aligned}
&W_r = W_o \left( X_{size} + \frac{1 - X_{size}}{X_{comp}} \right) \quad \text{,} 
&H_r = H_o \left( Y_{size} + \frac{1 - Y_{size}}{Y_{comp}} \right)
\end{aligned}
\label{eq: FR dimension}
\end{equation}
The dimensions of the FSC frame ($W_r, H_r$) are calculated using~\autoref{eq: FR dimension}.  The pseudocode for spatial compression along the x-axis is shown in Algorithm \autoref{algo:DFSC}, with a similar process for the y-axis. The boundary of the foveal region is determined by the compression ratios ($X_{comp}$, $Y_{comp}$) and the ROI size ($X_{size}$, $Y_{size}$). For every pixel ($i,j$) in the FSC frame, a mapping function identifies the corresponding locations ($i', j'$) in the game frame, as depicted in \autoref{fig:FR_DFSC}. The pixel sampling rate in the peripheral region is guided by $X_{comp}$ and $Y_{comp}$, with $c_p$ and $c_f$ as offset values for the mapping function. To ensure unnoticeable quality gradients, a smoothing function is applied to transition from lower to higher sampling rates as the distance from the center increases.

\begin{algorithm}[t!]
    \caption{Dynamic Foveated Spatial Compression (FSC)}
    \footnotesize
    \begin{algorithmic}[1]
        \State\textbf{Input:} $\text{frame}, W_r, W_o, X_o, X_{size}, X_{comp}$
        \State \textbf{Output:} FSC Frame
        \State $bound_{left} \gets \frac{2(1-X_{size})}{(X_{comp}-1)*X_{size}+1}*\frac{X_o}{W_o}$
        \State $bound_{right} \gets \frac{1-X_{size}}{X_{size}(X_{comp}-1)+1}*\frac{X_o-W_o}{W_o}+1$
        % \State \text{Initialize FSC frame}
        \For{$i$ from $0$ to $W_r$}
        \If{$i<bound_{left}$}\Comment{Peripheral region}
        \State $i' \gets ((X_{comp}-1)*X_{size}+1)*\frac{i}{W_r}*W_o$ 
        \State $i' \gets X_{comp}*i$
        \ElsIf{$i>bound_{right}$}\Comment{Peripheral region}
        \State $c_p \gets (1-X_{comp})*X_{size}$
        \State $i' \gets X_{comp}*i+W_o*c_p$
        \Else\Comment{Fovea region}
        \State $c_f \gets \frac{X_{comp}-1}{X_{comp}}*(1-X_{size})*(\frac{X_o}{W_o})$
        \State $i' \gets i + c_f*W_o$
        \EndIf
        \State \text{FSC frame}$[i] \gets \text{frame}[i']$
        \EndFor
        % \State\Return \text{FSC frame}
    \end{algorithmic}
    \label{algo:DFSC}
\end{algorithm}

\subsection{Gaze-driven Video Encoding}
\label{sec:video_encoding}
Prior works~\cite{foveatedEncodingForLargeResolutionDisplays,alhilal2024FovOptix,Illahi2020CGFVE} use fixed-size foveation areas or fixed quantization to lower bitrate demand in video streaming. In contrast, we model foveation and bitrate allocation as a Gaussian function driven by a foveation controller \(C\). This allows for dynamic adjustments of the foveation region, following bell-curve shape.

% \subsubsection{Eye Gaze Projection On Rendered Frame}
\textbf{ \circlednumber{1} Gaze Mapping: Screen to FSC Frame Coordinates.}  
FSC spatially compresses the game frame during the post-rendering stage to reduce peripheral resolution and generate the FSC frame. The compression is illustrated as pseudocode in Algorithm \ref{algo:DFSC}. Due to dimensional change, the eye gaze point within the FSC frame must be recalculated for accurate Foveated Video Encoding. To locate the gaze point $(X_r, Y_r)$ in the FSC frame, we use the foveated spatial decompression (FSD) as:
\begin{equation}
\begin{aligned}
&FSC(X_r, Y_r) = (X_o, Y_o)\quad \text{,} \quad
&X_r, Y_r = FSD(X_o, Y_o)
\label{eq: Gaze FR}
\end{aligned}
\end{equation}
The FSD function is the inverse of the FSC function, demonstrated in pseudocode in ~\autoref{app:sec4}.

\textbf{ \circlednumber{2} Video Encoder.}
\label{sec:encoder}
For low-latency video encoding, \sysname uses the Hardware-Accelerated Video Encoder (NVENC) \cite{nvidia_video_codec_sdk}, a hardware-based H.264/HEVC/AV1 video encoder for Nvidia GPU. Traditional frame encoding prioritizes the target bitrate as the primary parameter~\cite{nvidia_nvenc_2023} and allocates quantization parameters (QP) accordingly to optimize video quality while meeting to the bitrate target through multi-pass processing. In contrast, \sysname uses a single-pass encoding that prioritizes quality allocation aligned with the human visual system by enabling CONSTQP mode (\(QP_{const}\)) in the rate controller. In this mode, the encoder segments the entire frame into multiple macroblocks, each consisting of \(16\times16\) pixels. Each macroblock is associated with a Quantization Offset (QO), which is added to the \(QP_{const}\) to determine the Quantization Parameter (QP) for that macroblock. We create the QP encoding map by assigning QO to each macroblock within a frame. QP governs the quantity of spatial detail retained, with values ranging from 1 to 51. At its minimum, nearly all detail is retained (high level of detail). 
As QP increases, the level of detail decreases, leading to a reduction in bitrate, more distortion, and lower overall quality~\cite{nvidia_video_codec_sdk,pixeltools_rate_control}.
% \gap{write 2-3 sentences to state where you assign high QP and low QP with the frame, high level of detail (foveal region), and low level of detail in the periphery, guided by the gaze point obtained in equation 2. Then write, below details of the assignment of quality according to HVS }

\textbf{ \circlednumber{3} Non-uniform Quantization. } 
As shown in \autoref{fig:Gaze_projection}, \sysname generates a QP map for each frame, with QP values varying based on the distance from the gaze point. It assigns lower QP values to macroblocks near the gaze point, resulting in lower compression and higher visual quality. In contrast, higher QP values are assigned to macroblocks outside the foveal region, leading to greater compression and lower visual quality. This involves calculating the distance between the current macroblock and the macroblock containing the gaze point. Thus, we map the eye gaze point from the rendered frame coordinate system \((W_r,H_r)\) to the QP macroblocks coordinate system \((\frac{W_r}{16},\frac{H_r}{16})\). This allows us to locate the gaze macroblock containing the gaze point $(X_r,Y_r)$ in the QP map as follows:
\begin{equation}
    X_{QP} = \left\lceil \frac{X_r}{16} \right\rceil,
    Y_{QP} = \left\lceil \frac{Y_r}{16} \right\rceil 
    \label{eq: QP for X,Y}
\end{equation}
We compute each macroblock's Quantization Offset (QO) via a Gaussian function that models the foveation. Given the gaze macroblock\((X_{QP}, Y_{QP})\), the distance between any macroblock \((i,j)\) and gaze macroblock is \(\sqrt{(i-X_{QP})^2 + (j-Y_{QP})^2}\), hence the \(QO\) of this macroblock is calculated as follows:
\begin{equation}
    QO(i,j) = QO_{max} - QO_{max} \times \exp{(-\frac{(Distance(i,j))^2}{2C^2})}
    \label{eq: QO cal}
\end{equation}
where \(QO_{max} = QP_{max} - QP_{const}\) denotes the maximum value of QO, and \(C\) denotes the foveation controller. \(C\) facilitates extending and shrinking the foveation area according to varying network conditions, see \autoref{sec:integration} for details.  The process flow of gaze-driven video encoding is presented as a pseudocode algorithm in ~\autoref{app:sec5}

\subsection{Video Decoding and Client Side Rendering}
\label{sec:fvd_rr}
Video encoding takes a QP map with varying values. However, decoding using the H264 format does not require any additional information about the QP map. This eliminates the need to send the QP map to the client and ensures smooth video decoding. %After obtaining the decoded frame, we apply the $AADT'$ function to expand the FSC Frame back to the Game Frame for the reverse Foveated rendering. 
% \gap{Since the $AADT$ function is complex, the systematic reversal approach is applied to reverse each mathematical step and form the reversed AADT $(AADT')$}. 
% As shown in Figure\ref{fig:Reconstructed_frame}, the defined center region's fidelity of the reconstructed frame is preserved through the reverse foveated rendering process, while the remaining regions exhibit deterioration, particularly pronounced at the edge.
After decoding, we retrieve the FSC frame and apply FSD function to expand it back to the game frame for foveated spatial decompression. Notably, the fidelity of the foveal region is preserved, while it may drop in the outer regions, particularly at the edges. \autoref{fig:frames_over_arch} demonstrates video decoding and FSD at the client. A detailed visualization of the FSC and FSD effects is provided in ~\autoref{app:sec1}.
%As shown in \autoref{fig:Reconstructed_frame}, the fidelity of the foveal region is preserved, while the outer regions exhibit noticeable deterioration, particularly at the edges.  

\subsection{Network Monitoring}
\label{sec:network}
% To maintain low end-to-end latency through an unstable mobile network environment, the sending bitrate must adapt to fluctuations in network bandwidth. Motivated by \cite{GCCwebrtc}, we propose a strategy for tracking underlying network circumstances and detecting network congestion through analysing the transmission queuing delay gradient, which has been successfully utilised in numerous studies \cite{GCCwebrtc,queuingdelaygradient1,queuingdelaygradient2,queuingdelaygradient3,queuingdelaygradient4}. 
% The network monitoring model is established as the frame-level structure. For each frame, the server tracks server-side information. Once the client receives this frame, it transmits corresponding feedback packets containing client-side information back to the server for further processing. Through the analysis of frame-level information, the network monitoring model provides predictions regarding the current state of the network \((Normal, Underuse, Overuse, Timeout)\).
 To adjust the sending bitrate for fluctuations in available bandwidth, we monitor the one-way delay gradients of the transmission queue. This method effectively detects network congestion and bandwidth usage, as demonstrated in several studies \cite{GCCwebrtc,queuingdelaygradient1,queuingdelaygradient2,queuingdelaygradient3,queuingdelaygradient4}.

\textbf{ \circlednumber{1} Queuing Delay Gradient.} \sysname tracks the queuing delay gradient \((\nabla D)\) to assess bandwidth usage and detect congestions, thereby mitigating overshooting. To compute \(\nabla D\), \sysname maintains a moving window of data points \((t_i, D_i)\) for each frame, with a window size of 8 frames, where $t_i$ denotes the arrival time and \(D_i\) denotes the queuing delay of frame $i$ and computed as:
\begin{equation}
    D_i = (t_{arr}^{i} - t_{arr}^{i-1}) - (t_{send}^{i} - t_{send}^{i-1}) 
\end{equation} 
where \(t_{arr}^{i}\) and \(t_{send}^{i}\) denote the arrival time and sending time of frame i, respectively. We compute \(\nabla D_i\) by performing a linear regression on the data points.  The delay gradient for each frame is compared to a congestion threshold \(\gamma_{delay}\)  to determine the network state \(S_i\): if \(\nabla D > \gamma_{delay}\), it indicates bandwidth overuse; if \(\nabla D < -\gamma_{delay}\), it indicates bandwidth underuse; and when \(-\gamma_{delay} < \nabla D < \gamma_{delay}\), it indicates normal bandwidth usage.

\textbf{ \circlednumber{2} Feedback Timeout.}
The server detects network congestion using client feedback packets. Upon receiving new feedback, it estimates the network state \(S_i\). Due to fluctuations in mobile networks, feedback packets may be delayed or lost, negatively impacting adaptive video streaming performance~\cite{feedbackpkt,feedbackpkt1}. Thus, we compute the time interval between feedback packet arrivals as \(\delta t_{fd}^{i} = t_i - t_{fd}^{i-1}\), where \(t_i\) denotes the current timestamp, and \(t_{fd}^{i}\) and \(t_{fd}^{i-1}\) indicate the arrival time of the feedback packets for the \(i^{th}\) and \({i-1}^{th}\) frame. When \(\delta t_{fd}^{i} > \gamma_{fd}\), it indicates a timeout, signaling severe congestion that necessitates a significant reduction in sending bitrate for recovery.

\begin{figure}
    \centering
    \includegraphics[width=.6\linewidth]{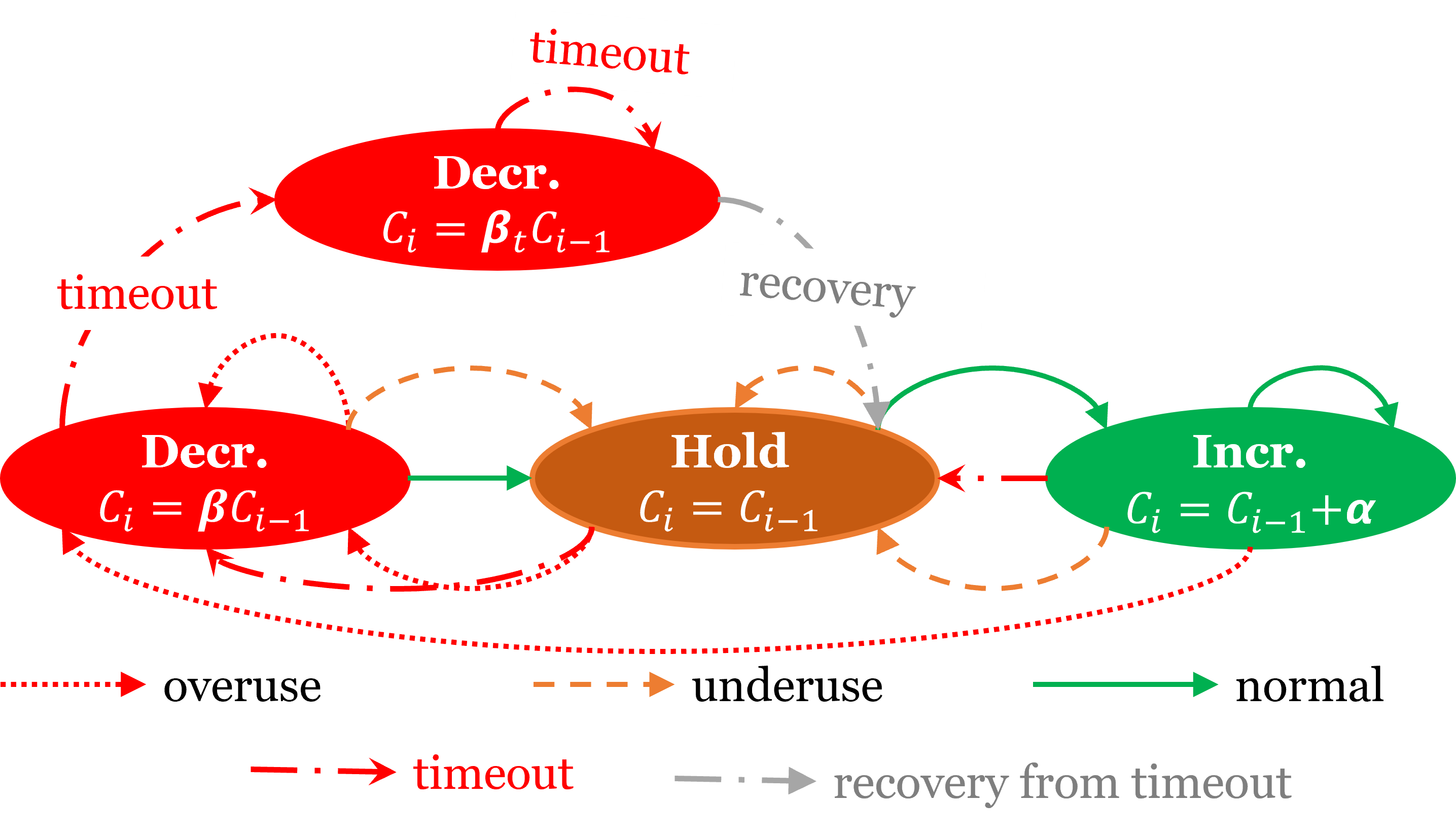}
    \caption{Quality allocation incorporating bandwidth usage: decrease, hold, or increase (eclipses) of the Foveation Controller \textbf{$C$} according to bandwidth usage (over/under-use, normal), or feedback timeout (severe congestion)  }
    \label{fig:controller_state_machine}
\end{figure}

\subsection{Gaze-contingent Rate Control}
\label{sec:integration}
Traditional rate control (ABR) algorithms provide a target bitrate as input to the video encoder~\cite{ABRSurvey,360ABRSurvey,GCCwebrtc,alvr}, which accordingly determines the optimal quantization (QP) uniformly. 
In contrast, FVE assigns QP non-uniformly, determining the sending bitrate. However, controlling the sending bitrate in FVE is challenging because it depends on the area of the foveal region and assigned QP values across the frame. Therefore, we propose a gaze-driven video encoding that takes a QP map as input after assigning QP values non-uniformly in compatibility with human vision perception. To assign the QP values in the QP map, we utilize a foveation controller finite state machine (\autoref{fig:controller_state_machine}) which associates the QP assignment with the network state \(S_i\). The controller helps to optimize visual quality and prevent network overshooting and undershooting by adjusting the foveation region according to the network state \(S_i\).

% \begin{figure}
%     \centering
%     \includegraphics[width=.9\linewidth]{figures/foveation_controller.pdf}
%     \caption{Quantization Offset (QO) across FoV regions and various values of foveation controller C, with central fovea span 5\degree, macular 18\degree, and periphery >18\degree \cite{degreesforfovea}.}
%     \label{fig:foveation_controller}
% \end{figure}

\textbf{ \circlednumber{1} Foveation Controller. }
As discussed in \autoref{sec:video_encoding}, \sysname computes QO and assigns QP values non-uniformly to each macroblock using the foveation controller \(C\), which adjusts to expand or shrink the foveation region, as illustrated in \autoref{fig:teaser}. Specifically, as  \(C\) increases, the standard deviation of the Gaussian curve increases, leading to a larger foveation region with a lower QP, while a decrease in  \(C\) narrows the curve, resulting in a smaller foveation region with a lower QP. \sysname utilizes a finite state machine to control \(C\), adjusting the transmission bitrate in response to fluctuations in network bandwidth. The C value ranges between predefined $C_{min}$ and $C_{max}$. The configuration of this range is discussed in the implementation (Section~\ref{sec:impl}). Notably, configuring \(C_{min}\) and \(C_{max}\) depends on the frame resolution. 

\textbf{ \circlednumber{2} Additive Increase and Multiplicative Decrease (AIMD). }
\sysname employs AIMD~\cite{GCCwebrtc,NADA}, utilizing multiplicative decrease to recover from network congestion while maintaining low end-to-end latency, and additive increase when available bandwidth rises. Specifically, as shown in ~\autoref{fig:controller_state_machine}, \sysname increases \(C\) by an incremental factor of \(\alpha = 0.2\) when network throughput improves, and decreases \(C\) by multiplying with the decrement factor \((\beta)\). If a feedback timeout occurs, a sharper timeout multiplier \(\beta_t\) is applied, where \(\beta = 0.9, \beta_t = 0.85\).

\section{Implementation} 
\label{sec:impl}
\sysname includes server and client applications (see \autoref{fig:architecture}). The computationally intensive tasks are primarily handled on the server to reduce the load on the client. To create a VR headset and game-independent system, we implement \sysname on the ALVR base code\cite{alvr}.

% Client Side implementation
\subsection{Client}
We implement the client as an Android application and install it on Meta Quest Pro, with a target frame rate set at 72 Hz. %We use Rust and multithreading to implement the pipeline, building on ALVR codebase.

\textbf{Data Collection.} We leverage OpenXR API~\cite{OpenXR} to collect the raw data of eye gaze, specifically the gaze orientation in 3D space \((f_{yaw},f_{pitch})\), at a frequency of 200 Hz, which is nearly three times the desired frame rate (72 FPS). Since VR headsets display a frame for each eye (left-eye and right-eye frame)~\cite{vrframecomposition}, we enable gaze orientation tracking for each eye.
We use Rust language to implement network monitoring and communication between the client and the server. The client transmits the eye gaze and motion data to the server over a UDP socket at 200 Hz, while transmitting the network data at frequency of frame receiving rate.

\textbf{Video Decode.} The client uses Android MediaCodec~\cite{android_media_codec} to decode frame packets. As mentioned in~\autoref{sec:fvd_rr}, decoding is straightforward and requires no additional input, despite utilizing spatially variable QP values to create a QP map for encoding. The frames are then stored in a FIFO queue, awaiting processing by the decompression thread.

\textbf{Foveated Spatial Decompression.} We implement FSD using OpenGL shader language (GLSL) based on the OpenGL library~\cite{OpenGL}. This retrieves the game frame from the FSC frame, recovering the original game frame's resolution. In our implementation of~\autoref{sec:gaze_rendering}, we fix  \((X_{size} = 0.45, Y_{size} = 0.4)\) and \((X_{comp} = 4, Y_{comp} = 5)\), while the gaze point \((X_o, Y_o)\) remains variable, extracted from the header of received video packet. After FSD, the frame is stored in a FIFO queue, awaiting display, following the conventional protocol of VR frame lifecycle \cite{meta_horizon_vr_frames}.

% Server Side implementation
\subsection{Server}
Similar to ALVR, we develop the server as a Windows application, with modifications to the rendering and video encoding in the streaming pipeline. %We implement the pipeline based on multithreading using Rust and build upon the ALVR codebase.

\textbf{Foveated Spatial Compression.}
We implement a pixel shader using High-level Shading Language (HLSL) within Direct3D 11~\cite{Stevewhims} because the server operates as a Windows application. The shader is applied to the original game frame, consisting of the left and right-eye frames. After obtaining the gaze data, the shader is reinitiated to map the new gaze data to each eye frame.

\textbf{Quality Manager.}
We implement this unit using Rust to monitor the network performance (Section \ref{sec:network}), and control the bitrate based on gaze data (Section \ref{sec:integration}). Upon receiving a feedback packet, the manager synchronizes the server data with the client feedback data using a timestamp recorded during the acquisition of the motion data. Using the network state and gaze mapped point, the manager adjusts the value of the foveation controller \(C\) (discussed in Section~\ref{sec:sysdesign}). We configure the  \(C\) range to [\(C_{min} = 6\), \(C_{max} = 120\)]. This ensures that the QP values of the macroblocks within the central fovea region \(5\degree\) and the most distant areas, respectively, do not grow larger than 23, where QP=23 is the default configuration for the recommended quality in H.264 CRF mode~\cite{FFmpegH264}.

\textbf{Video Encode.}
We implement the foveated video encoder in C++. We leverage the server's Nvida GPU to perform hardware-accelerated video encoding in H264 format through the NVIDIA Video Codec SDK~\cite{nvidia_video_codec_sdk}. Following the recommended NVENC configurations for low-latency applications~\cite{nvidia_nvenc_2023}, we configure the NVENC encoder with an infinite GOP length and low-latency tuning parameters. We employ the CONSTQP rate control mode and define \(QP_{const} = 11\), as prior research indicates that a QP value lower than 11 does not produce any discernible improvement and rather escalates the needed bandwidth. % Prior to performing video encoding on fresh frames, the encoder retrieves the most recent value of the foveation controller \(C\) and the gaze macroblock \((X_{qp},Y_{qp})\) from the quality manager component, subsequently calculating QO for each macroblock utilising the Gaussian function illustrated in \autoref{sec:video_encoding}. 
We leverage the emphasis map feature in the NVENCODE API~\cite{EmphasisMAP_nvenc} to apply varying quality levels at the macroblock level. This allows us to use the generated QP map when encoding frames. Encoded frame packets are sent to the client via a UDP socket, including headers with the synchronization key (acquisition timestamp) and the mapped gaze center \(X_o, Y_o\). We also record the sending timestamp \(t_{send}\) to calculate the client-to-server delay.

% We implement the server as a Windows application. We implement the foveated rendering and foveated video encoder in C++, while networking and network monitoring model in Rust. The server's implementation follows to the ALVR structure \cite{alvr}, and by integrating each component, we establish our own streaming pipeline. We build the pixel shader to execute foveated rendering utilising the latest eye gaze data for the original composite frame, composed of the left-eye and right-eye frames. This procedure is illustrated in Section: \ref{sec:sysdesign} and replicated for both eyes. We leverage the server's GPU to perform hardware-accelerated video encoding in H264 format through the NVIDIA Video Codec SDK \cite{nvidia_video_codec_sdk}. In accordance with the suggested NVENC configurations for low-latency applications \cite{nvidia_nvenc_2023}, we set an infinite GOP length and low latency tuning parameters for the NVENC encoder. Furthermore, as outlined in Section: \ref{subsubsec:encoder}, we employ the CONSTQP rate control mode and define \(QP_{const} = 11\), as prior research indicates that a QP value lower than 11 does not produce any discernible improvement and rather escalates the needed bandwidth. Utilising the eye gaze projection results in the rendered frame and the QO value for each macroblock, we apply the emphasis map feature in the NVENC API for video encoding. Simultaneously, upon the receipt of a new feedback packet, the network monitoring model will be updated. 

\section{Experiment Setup} 
\label{sec:setup}

This section covers the experimental testbed, the network traces used to emulate real-life mobile cloud gaming, and the evaluation benchmarks.

\subsection{Experimental Testbed}
\label{sec:testbed}

\begin{figure}[H]
    \centering
    \includegraphics[width=.6\linewidth]{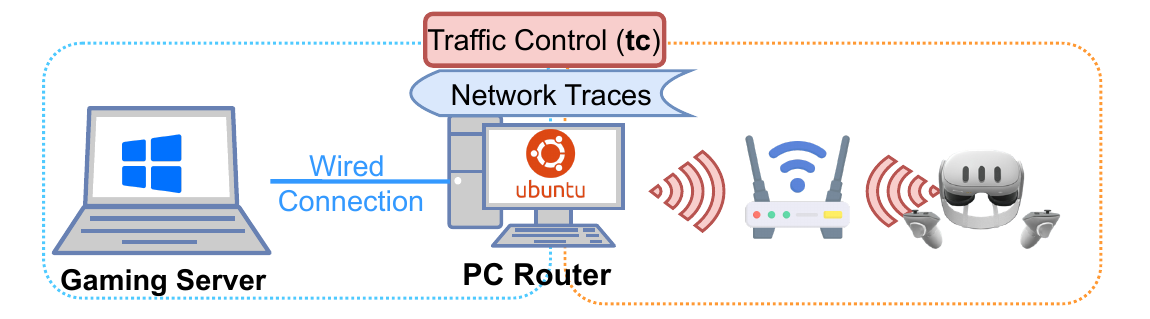}
    % \vspace{-.3cm}
    \caption{Cloud gaming emulator with a PC router incorporating 5G mobile network traces using traffic control (tc), emulating VR headset to cloud network.}
    \label{fig:testbed}
\end{figure}

As illustrated in~\autoref{fig:testbed}, we first design a physical testbed to create a realistic emulation of cloud gaming over variable network conditions. We configure a Linux PC with WiFi and Ethernet interfaces (Ubuntu 22.04.3 LTS, Intel(R) Core(TM) i9-13900K @ 5.80 GHZ) to function as a router between wired and wireless networks. This router relays traffic data between an Oculus Quest Pro (Android 12, Octa-core Kryo 585, Adreno 650) over a wireless network and a Windows PC (Windows 11 24H2, Intel(R) Core(TM) i9-14900KF @ 6.0 GHZ, NVIDIA GeForce RTX 4090 GPU, Driver Version: 565.90) over Ethernet to serve as a cloud server. We run a bash script on the PC router to emulate mobile network connectivity for the Oculus. The script utilizes Linux tc~\footnote{\url{https://man7.org/linux/man-pages/man8/tc.8.html}} to impose a variation in network throughput based on 5G traces.% a queuing delay (20ms) for the arrival packets before departing on the outgoing WiFi interface.

\subsection{Mobile Network Traces}
\label{sec:traces}
We leverage a corpus of \textbf{\textit{5G mobile network traces}} to emulate the real-world network between the player's VR headset and the cloud. The traces are collected from a major Irish mobile operator, and generated over video streaming and file download. We use the traces. Specifically, they are generated during continuous large file downloads, streaming of video content from Netflix service provider, and streaming of video content from Amazon Prime service provider~\cite{raca2020beyond}. %VR gamers typically play VR games at home or in stationary conditions. Therefore, we choose the traces of static patterns (out of driving and static) to reflect the typical game-playing conditions. 
We filter the traces to retain only records with a throughput exceeding 10 Mb/s, as some baselines fail at bandwidths below this threshold. 

\subsection{Baselines}
\label{sec:baselines}

\begin{table}[!t]
    \centering
    \scriptsize
    \caption{Evaluation Benchmarks}
    \label{tab:baselines}
\begin{tabular}{p{1.8cm}|p{2.8cm}|p{8.6cm}}
% \toprule
\textbf{Protocol} & \textbf{Rate Control} &  \textbf{Description}  \\
\midrule

\textbf{ALVR}~\cite{alvr} Open-source & frame interval and throughput & \textbf{VR gaming over WiFi:} adjust bitrate according to frame interval and throughput, and reduce computation via FFR\\

\midrule
\textbf{GCC}~\cite{GCCwebrtc}   Industry Standard & latency, bandwidth utilization, and loss& \textbf{Google Standard:} adjust bitrate according to delay gradients, bandwidth usage and packet loss\\

\midrule
\textbf{CGFVE}~\cite{Illahi2020CGFVE} Research SoTA  & foveation with TCP &   Non-uniformly assign quality in video encoding based on real-time gaze to reduce bandwidth demand, designed for 2D PC-based environments.\\

\midrule
\textbf{FovOptix}~\cite{alhilal2024FovOptix}  Research SoTA  & fixed foveation, latency, and bandwidth usage & Non-uniformly assign quality in fixed FoV regions based on latency and bandwidth usage, and reduce computation load via fixed foveated rendering.\\

\midrule
\textbf{\sysname} & gaze point, latency and bandwidth usage  &  Non-uniformly assign resolution and quality based on real-time gaze, delay gradients, and bandwidth usage\\

\end{tabular}
\end{table}

Table~\ref{tab:baselines} summarizes the benchmarks, including Google Congestion Control (GCC) which is developed by Google as an industry standard\cite{GCCwebrtc}, open-source VR gaming over WiFi (ALVR)\cite{alvr}, and state-of-the-art research methods (CGFVE~\cite{Illahi2020CGFVE} and FovOptix~\cite{alhilal2024FovOptix}).

\begin{figure}[!t]
    \centering
    \begin{subfigure}[b]{.67\textwidth}
        \includegraphics[width=1.03\linewidth]{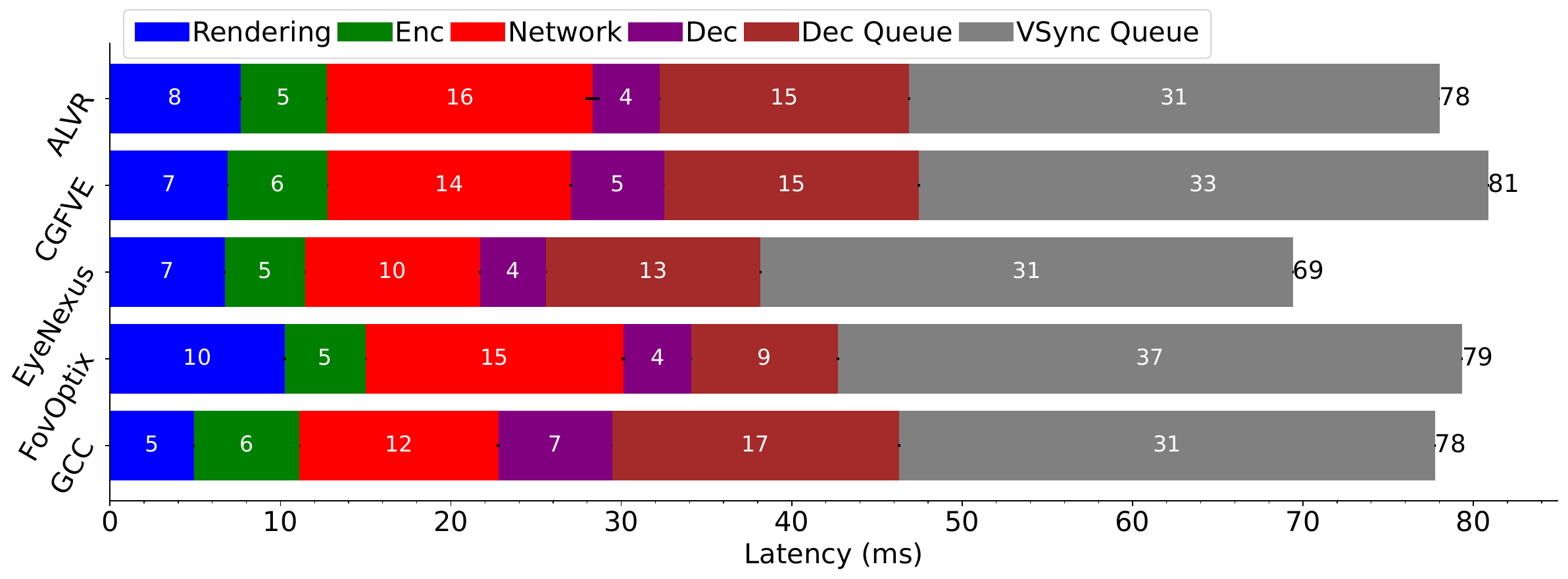}
        \caption{End-to-end Latency Decomposition over WiFi}
    \label{fig:wifi_latency_stackbar}
    \end{subfigure}
    \hfill
     \begin{subfigure}[b]{.3\textwidth}
        \includegraphics[width=1.03\linewidth]{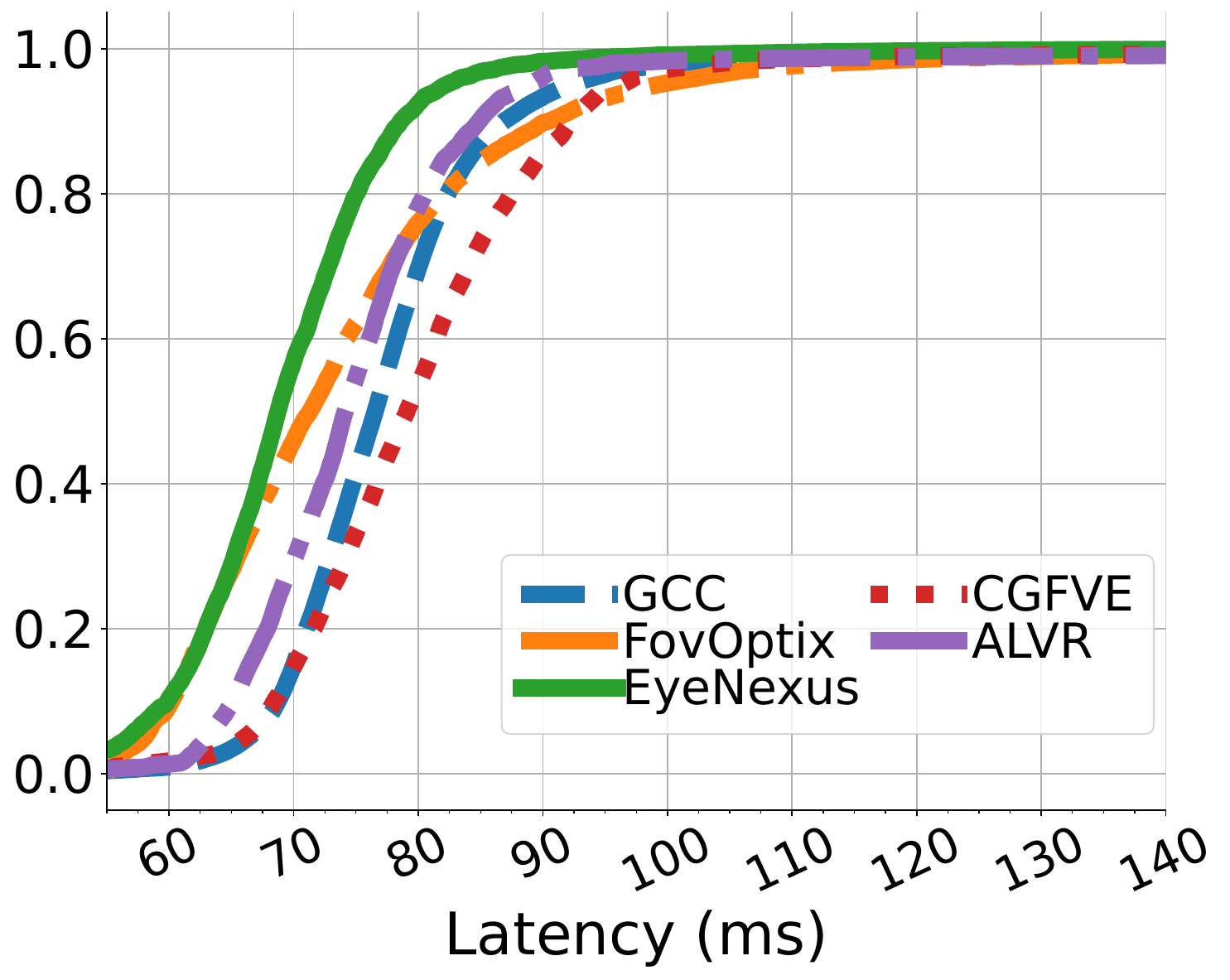}
        \caption{CDF of MTP Latency (WiFi)}
    \label{fig:WiFi_latency_CDF}
    \end{subfigure}
    \begin{subfigure}[b]{.67\textwidth}
        \includegraphics[width=1.03\linewidth]{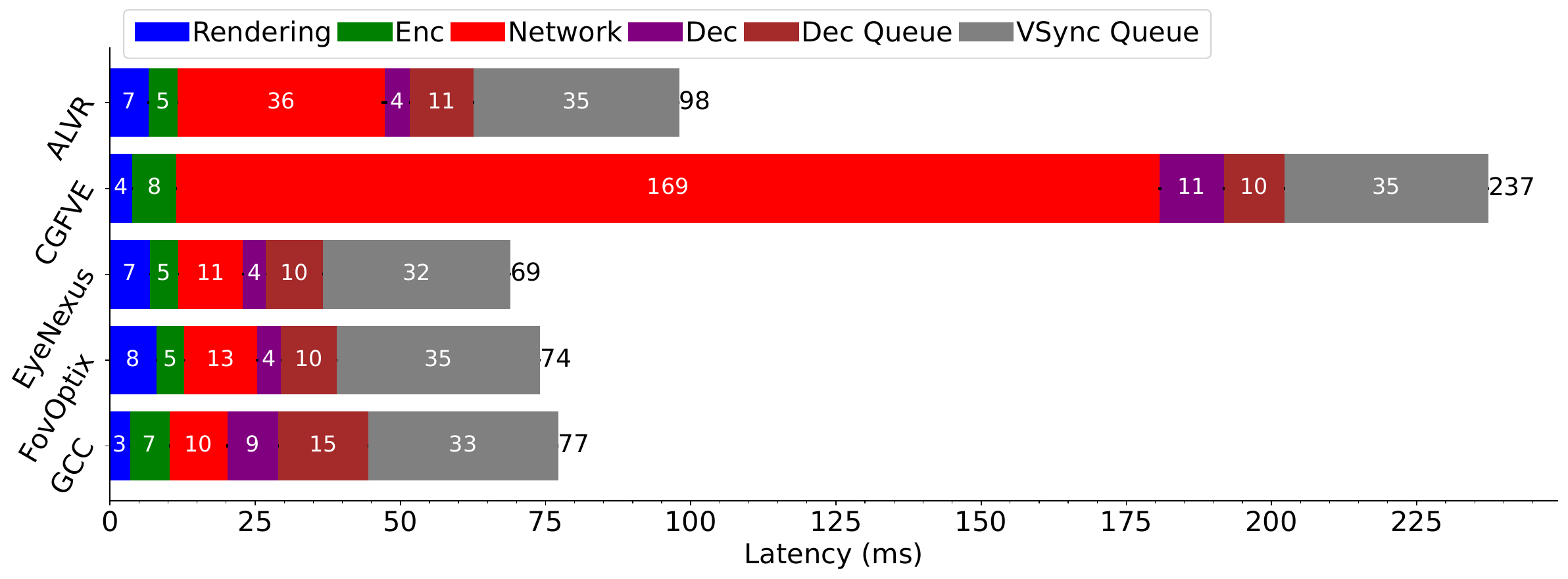}
        \caption{End-to-end Latency Decomposition over Mobile Network}
    \label{fig:mobile_latency_stackbar}
    \end{subfigure}
    \hfill
     \begin{subfigure}[b]{.3\textwidth}
        \includegraphics[width=1.03\linewidth]{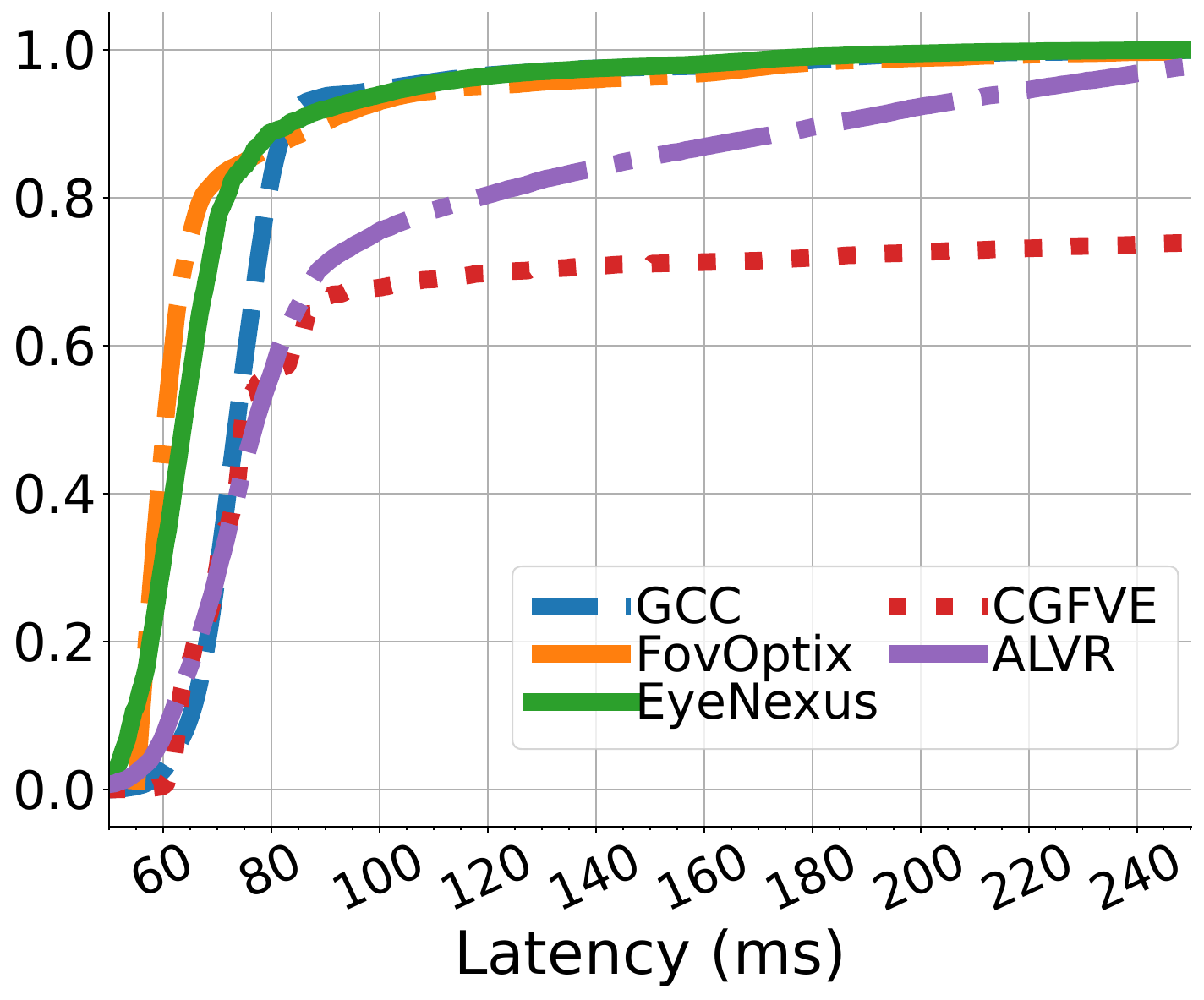}
        \caption{CDF of MTP Latency (Mobile)}
    \label{fig:mobile_latency_CDF}
    \end{subfigure}
    \vspace{-.2cm}
    \caption{Latency decomposition and cumulative distribution of MTP latency over WiFi and mobile network.}
    \label{fig:wifi_mobile_latency}
\end{figure}

\section{Performance Evaluation}
\label{sec:eval}
There are two main setups for remote-rendered VR games: 1) a VR headset connected to a high-end gaming server via WiFi (local rendering offloading), and 2) a cloud-based gaming server accessed by the VR headset through mobile networks (cloud rendering offloading). In this section, we evaluate the performance of \sysname over the two setups. 

\subsection{End-to-end and Network Latency}

\textbf{Performance over Wireless Network.} \autoref{fig:wifi_latency_stackbar} illustrates the latency decomposition and cumulative distribution of end-to-end latency for all streaming protocols over a real-world wireless network (university WiFi—eduroam). \sysname leads all baseline systems with the lowest end-to-end latency (mean 69.4 ms), followed by GCC (mean 77.7 ms), ALVR (mean 78 ms), FovOptix (mean 79.3 ms), and CGFVE (mean 80.9 ms).  Additionally, \sysname's foveated video encoding substantially lowers network latency, establishing it as superior to the benchmarks. CGFVE is designed for PC-based gaming and operates at a lower resolution than VR. Additionally, GCC-based WebRTC (Google Standard) and CGFVE utilize uniform rendering, increasing the video encoding time. Specifically, larger rendered frames require longer encoding times by the video encoder.

% Although \sysname exhibits the highest Rendering latency (mean 3.4 ms) due to its dynamic gaze-guided foveated rendering (FR) component, this design significantly reduces Video Encoding and Decoding latency.

We present the CDF of MTP latency for the compared protocols in \autoref{fig:WiFi_latency_CDF}. 
% percentile 1% - 99%
\sysname shows the lowest MTP latency (mean 69.4 ms) with minimal variation (51.7-98.3 ms) for 1\% to 99\% of playtime. GCC has a higher mean latency of 77.7 ms with the second lowest variation (60.4-109.8 ms), followed closely by ALVR with a mean of 78 ms and a variation range of (58.2-131.8 ms). FovOptix follows with a mean latency of 79.3 ms and a broader variation (55.6-131.1 ms), while CGFVE has a mean latency of 80.9 ms and a variation range of (55.8-123.9 ms) for 1\% to 99\% of playtime. 
% percentile 80% 
\sysname keeps latency below 75 ms for 80\% of the playtime. In contrast, ALVR, FovOptix, GCC, and CGFVE show higher MTP latency with greater variation, maintaining latencies below 80.5, 82, 82.3, and 87.5 ms for 80\% of playtime, respectively. 

\textbf{Performance over Mobile Network.} \autoref{fig:mobile_latency_stackbar} shows the latency decomposition and cumulative distribution of end-to-end latency for all streaming protocols over a realistic emulation of 5G mobile networks. \sysname exhibits the lowest latency, averaging 69 ms, while FovOptix achieves the second lowest latency at 74 ms. GCC follows with an average latency of 77 ms, which is attributed to its design for real-time communication, which favors latency over visual quality, as indicated by its lowest network latency among the benchmarks. Conversely, ALVR performs poorly under variable mobile network conditions, resulting in a higher average latency of 98 ms. Although CGFVE incorporates FVE to reduce bandwidth demand, it does not adapt quality allocation to the available bandwidth, leading to a significant increase in network latency (mean 169 ms) and the highest end-to-end latency, with a mean of 237 ms.

We present the CDF of MTP latency for the compared protocols in \autoref{fig:mobile_latency_CDF}. \sysname exhibits the lowest MTP latency (mean 69 ms) with minimal variation (48.7-178.8 ms) for 1\% to 99\% of playtime. Although FovOptix exhibits the second lowest latency (74 ms), its variation (55.1-207.9 ms) is broader than GCC's variations (57.3-186.9 ms) for 1\% to 99\% of playtime. ALVR exhibits a broader variation (51.6-264.7 ms), while CGFVE performs poorly, showing the highest MTP latency (mean 237 ms) and the widest variation (60.6-1151.9 ms) for 1\% to 99\% of playtime. Notably, CGFVE's MTP latency exceeds 1 second for 4\% of gameplay (the upper 4\% tail) due to inflated network latency. This latency inflation arises from CGFVE's quality adjustment, which sets a fixed QP based on a predefined rule, overlooking variations in available bandwidth under mobile network conditions. 
% percentile 90% 
\sysname and GCC maintain latency below 83.3 ms, while FovOptix keeps latency below 89.5 ms for 90\% of playtime. In contrast, ALVR and CGFVE experience significant increases in latency, reaching 183.4 ms and 749.8 ms, respectively, for 90\% of playtime. The significant rise in CGFVE's latency is notably due to inflated network latency resulting from rigidity in quality allocation.

\begin{figure}
    \centering
    \begin{subfigure}[t]{0.49\linewidth}
        \centering
        \includegraphics[width=.9\linewidth]{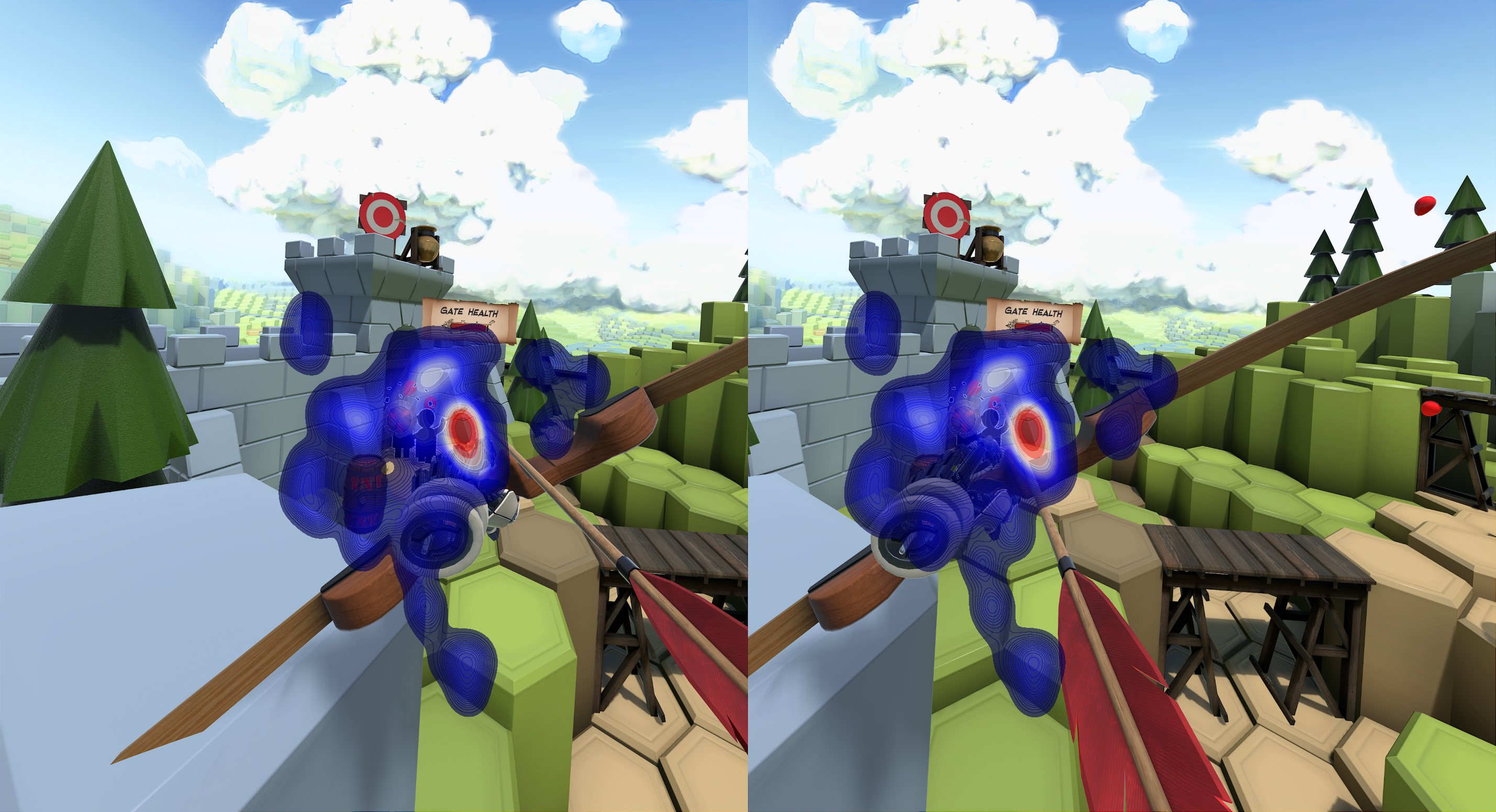}
        \caption{Heatmap of participants' gaze points in FPS game}
        \label{fig:heatmap_fps}
    \end{subfigure}
    \hfill
    \begin{subfigure}[t]{0.49\linewidth}
        \centering
        \includegraphics[width=.9\linewidth]{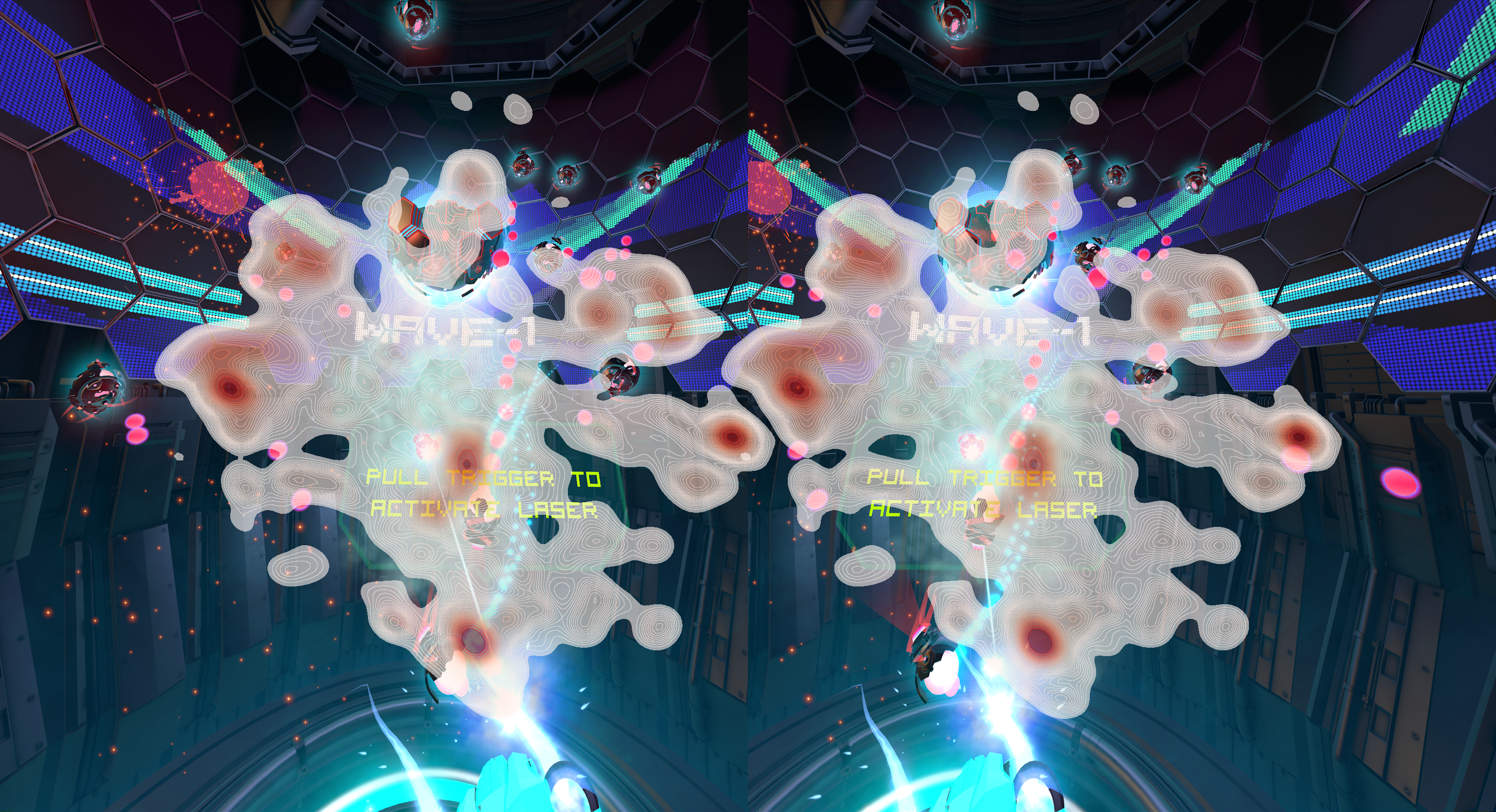}
        \caption{Heatmap of participants' gaze points in TPS game}
        \label{fig:heatmap_tps}
    \end{subfigure}
    % \vspace{-.2cm}
    \caption{Variations of visual attention (gaze focus) between (a) First-person shooting (FPS) game, and (b) Third-person shooting (TPS) game. Heatmaps of the eye gaze data, collected during user experiments.}
    \label{fig:heatmap_comparison}
\end{figure}

% Figure: Performance over Wireless/Mobile Network.
\begin{figure}[!t]
    \centering
    \begin{subfigure}[b]{.24\textwidth}
        \includegraphics[width=1.05\linewidth]{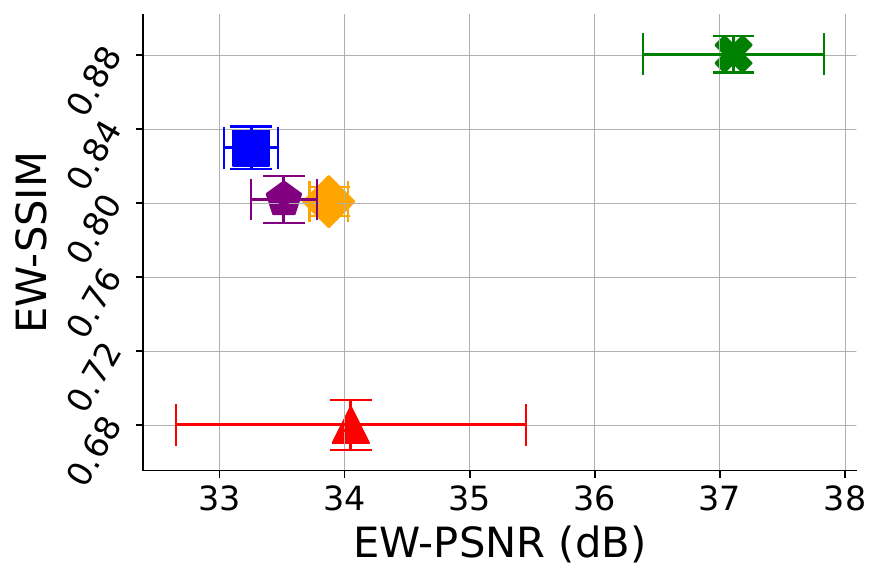}
        \caption{TPS game over WiFi }
    \label{fig:WiFi_ewssim_ewpsnr_TPSgame}
    \end{subfigure}
     \begin{subfigure}[b]{.24\textwidth}
        \includegraphics[width=1.05\linewidth]{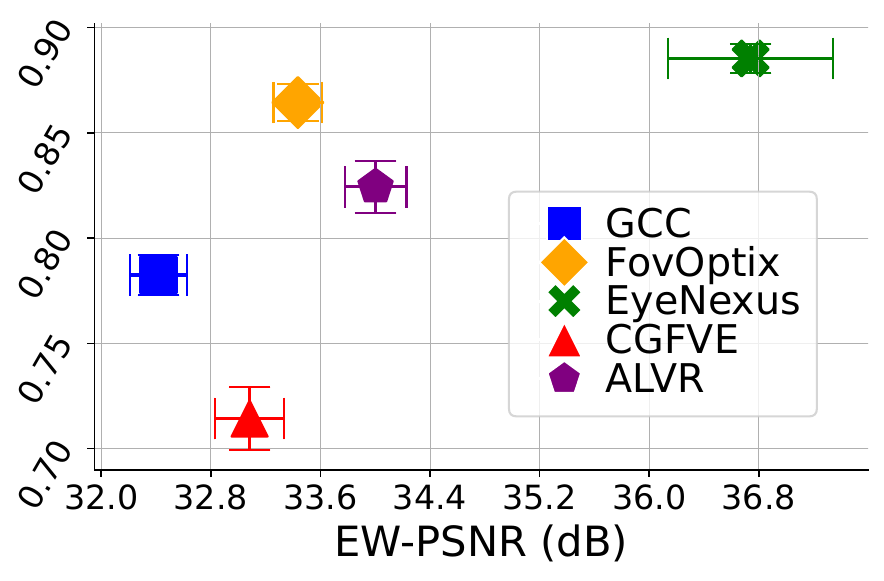}
        \caption{TPS game over mobile}
    \label{fig:mobile_ewssim_ewpsnr_TPSgame}
    \end{subfigure}
    \begin{subfigure}[b]{.24\textwidth}
        \includegraphics[width=1.05\linewidth]{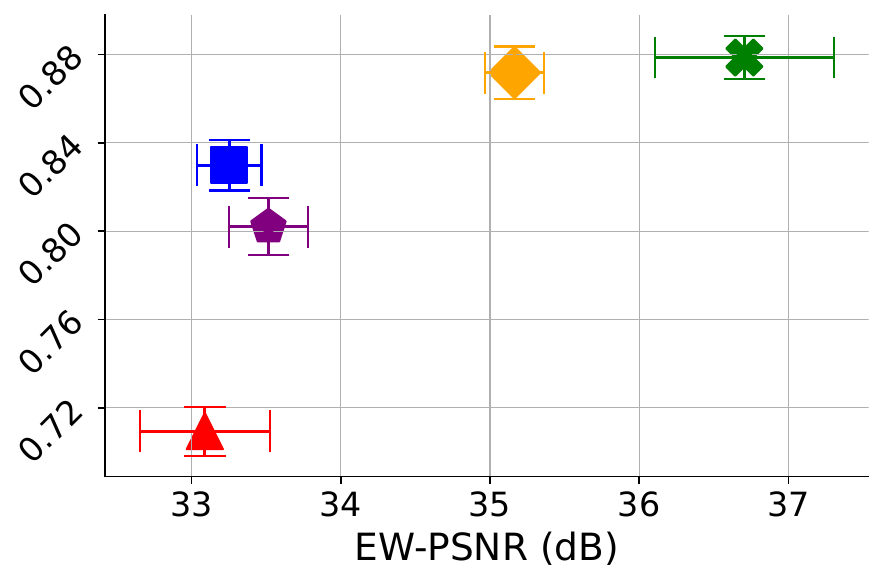}
        \caption{FPS game over WiFi }
    \label{fig:WiFi_ewssim_ewpsnr_FPSgame}
    \end{subfigure}
     \begin{subfigure}[b]{.24\textwidth}
        \includegraphics[width=1.05\linewidth]{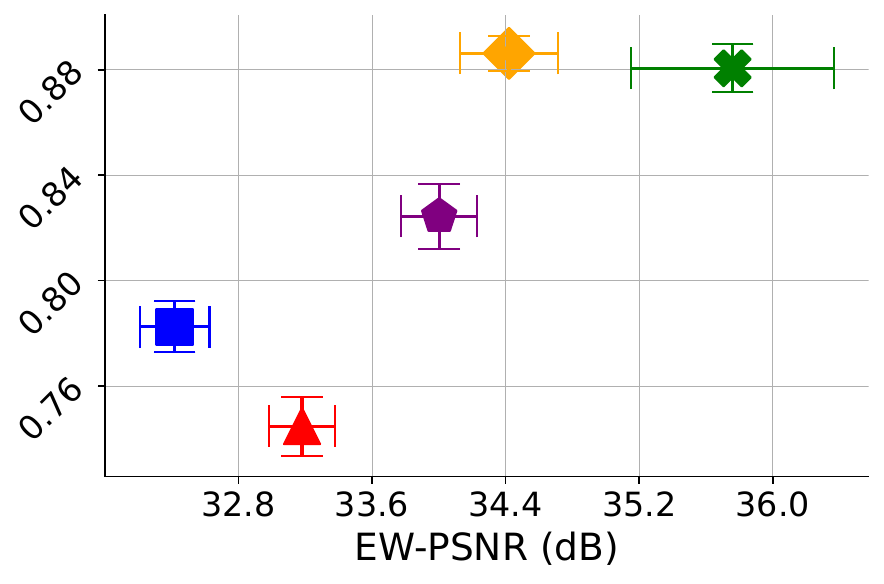}
        \caption{FPS game over mobile}
    \label{fig:mobile_ewssim_ewpsnr_FPSgame}
    \end{subfigure}
    \caption{Visual quality (EWSSIM and EWPSNR) of delivered TPS and FPS game frames over (a) WiFi network, and (b) emulation of 5G mobile network.}
    \label{fig:ewssim_ewpsnr_TPSgame}
\end{figure}

% Visual Quality (Eye weighted PSNR/SSIM) 
\subsection{Visual Quality (VQ)} 
We assess the visual quality (VQ) of \sysname using two metrics: EWSSIM~\cite{fassim2011foveation} and EWPSNR~\cite{li2011visual}. Both metrics are designed to evaluate foveated frames and are used in related work~\cite{Theia,Illahi2020CGFVE}. They assign weights to image pixels according to the actual gaze direction, with the weights gradually diminishing from the center (fovea) to the edges (periphery), following the human visual system (HVS). \autoref{fig:ewssim_ewpsnr_TPSgame} illustrates the VQ measured by EWSSIM (y-axis) and EWPSNR (x-axis) for both WiFi (\autoref{fig:WiFi_ewssim_ewpsnr_TPSgame} and \ref{fig:WiFi_ewssim_ewpsnr_FPSgame}) and mobile network setups (\autoref{fig:mobile_ewssim_ewpsnr_TPSgame} and \ref{fig:mobile_ewssim_ewpsnr_FPSgame}). \sysname achieves the highest VQ in both metrics, while CGFVE presents the lowest EWSSIM over WiFi and mobile networks. Despite GCC using uniform rendering, its adaptability to network variations secures it in third place. In contrast, CGFVE and ALVR have the lowest VQ, with CGFVE being the worst. This is due to CGFVE's rigid inter-frame quality allocation, which fails to adapt over time, while ALVR's uniform encoding relies on a simple ABR algorithm that struggles under network fluctuations.

\subsection{Performance Over Higher Resolutions.}
\label{sec:high_res}

\begin{figure}
    \centering
    \begin{subfigure}[b]{.55\textwidth}
    \includegraphics[width=1.12\linewidth]{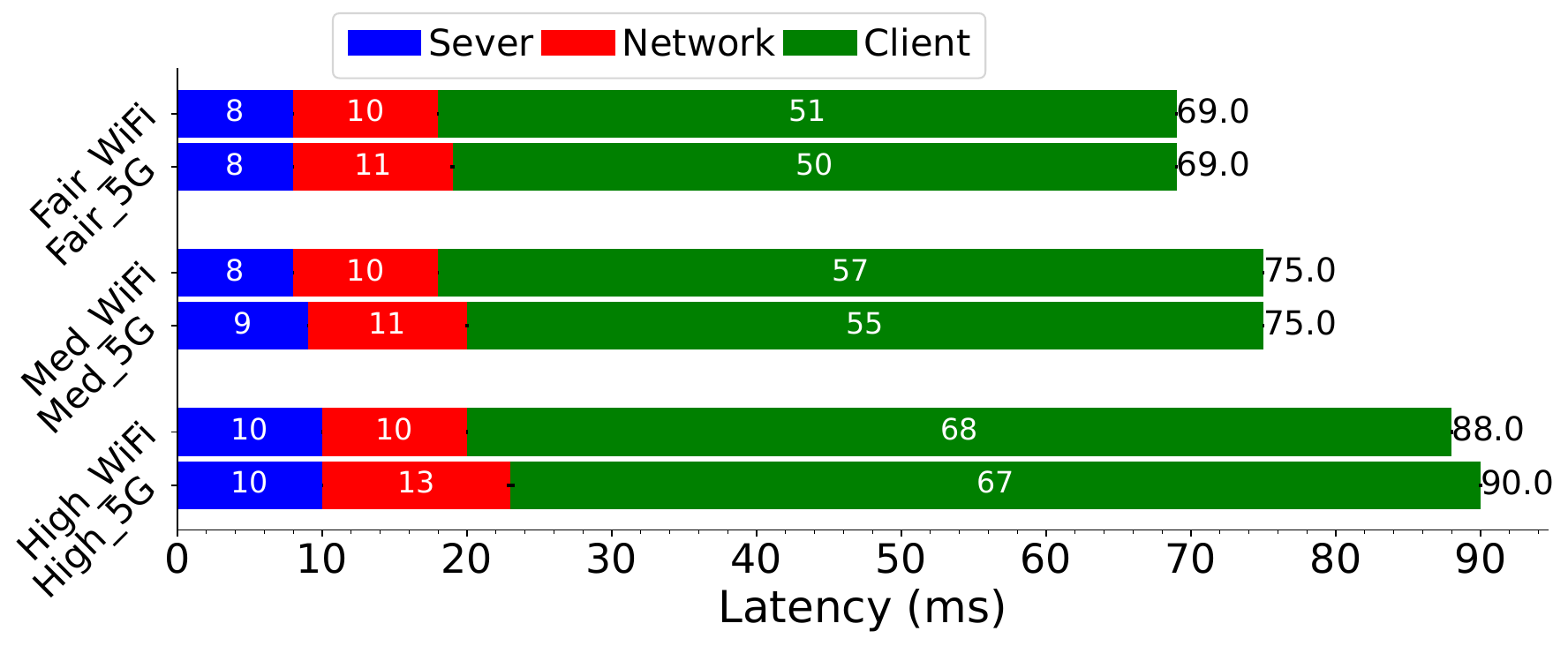}
    \caption{Latency decomposition of MTP latency}
    \label{fig:high_res_latency}
    \end{subfigure}
    \hfill
    \begin{subfigure}[b]{.38\textwidth}
        \includegraphics[width=.95\linewidth]{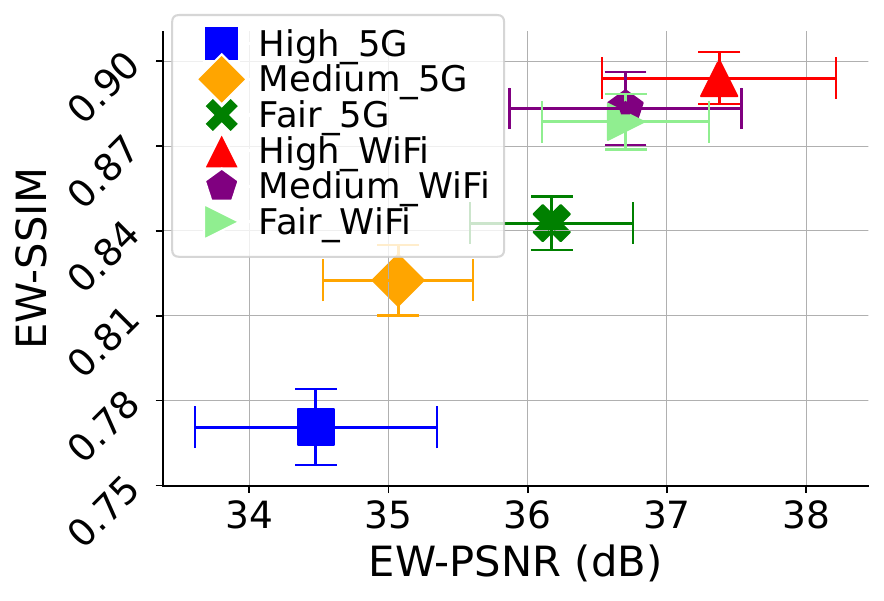}
        \caption{Visual Quality (VQ)}
    \label{fig:high_res_vq}
    \end{subfigure}
    % \vspace{-.2cm}
    \caption{MTP Latency and Visual Quality (EWSSIM and EWPSNR) for increasing resolutions over WiFi and Mobile Network setups, with resolutions [High \((5184\times2848)\), Medium \((4288\times2336)\), Fair \((3712\times2016)\)]}
    \label{fig:high_res_latency_vq}
\end{figure}

Hardware-accelerated video codecs support a maximum resolution of 4K (\(4096 \times 4096\)) according to Nvidia\footnote{\url{https://developer.nvidia.com/video-codec-sdk}}. FSC enables \sysname to accommodate higher resolutions and scale up with increasing resolution. To understand the scalability level, we examine \sysname end-to-end latency and visual quality across three resolutions: High (\(5184\times2848\)), Medium (\(4288\times2336\)), and Fair (\(3712\times2016\)).\\
\textbf{Latency over WiFi and Mobile Network.} \autoref{fig:high_res_latency} shows the breakdown of end-to-end latency for game frames at three resolutions. %For Fair resolution, the average latency is 69 ms on WiFi and mobile networks. For Medium resolution, it is 74 ms on WiFi and 75 ms on mobile networks. For High resolution, \sysname exhibits an average latency of 87 ms on WiFi and 89 ms on mobile networks. The MTP latency indicates an upward trend as resolution increases; however, \autoref{fig:high_res_latency} illustrates the latency decomposition into server, network and client latency, indicating that the additional latency mainly originates from the client side. 
 Average latency increases linearly with resolution: 69 ms for Fair, 74 ms for Medium, and 87 ms for High on WiFi, with corresponding values of 69 ms, 75 ms, and 89 ms for mobile networks. This trend demonstrates scalability as resolution increases. The latency decomposition reveals that the additional latency primarily arises from the client side, which can be attributed to the limited resources of the VR headset (Meta Quest Pro). These constraints hinder the processing capability required for higher resolutions, leading to increased latency as demands on the client resources grow.
 %Consequently, we ascertain that the escalating latency is attributable to the hardware constraints of Meta Quest Pro when processing high-resolution images. 
 %To substantiate this, we perform an ablation study on the latest VR headset (Meta Quest 3\footnote{{\url{https://www.meta.com/quest/compare/}}}), which possesses superior capabilities compared to the Meta Quest Pro.  The corresponding results validate our conclusion, as presented in \autoref{app:Ablation study}.
 To support this, we conduct an ablation study using the Meta Quest 3, as it offers advanced capabilities relevant to our analysis. The results of this study back up our conclusion, as detailed in ~\autoref{app:sec3}.\\
 \textbf{Visual Quality over Mobile and WiFi Networks.} As illustrated in \autoref{fig:high_res_vq}, \sysname shows a marginal drop in EWPSNR and EWSSIM for higher resolutions. This is attributed to the fluctuation in the mobile network throughput that leads to frequent multiplicative decreases in sending bitrate. Specifically, the average EWPSNR and EWSSIM values are 36.2 dB and 0.85 for Fair resolution, 35.1 dB and 0.82 for Medium, and 34.5 dB and 0.77 for High resolution. In contrast to the mobile network setup, \sysname shows a consistent increase in EWPSNR and EWSSIM for higher resolutions in the WiFi setup. Specifically, \sysname attains an EWPSNR with average values of 37.3, 36.7, and 36.7 and an EWSSIM with means of 0.879, 0.883, and 0.894 for Fair, Medium, and High resolutions, respectively. We attribute this to the higher stability of WiFi throughput. 
The superior visual quality for higher resolutions demonstrates \sysname's capability to further improve the user's QoE over stable networks with higher bandwidth.

\subsection{Ablation Study}
%Figure: ablation study
\begin{figure}
    \centering
    \begin{subfigure}[t]{0.49\textwidth}
        \centering
        \includegraphics[width=1.2\linewidth]{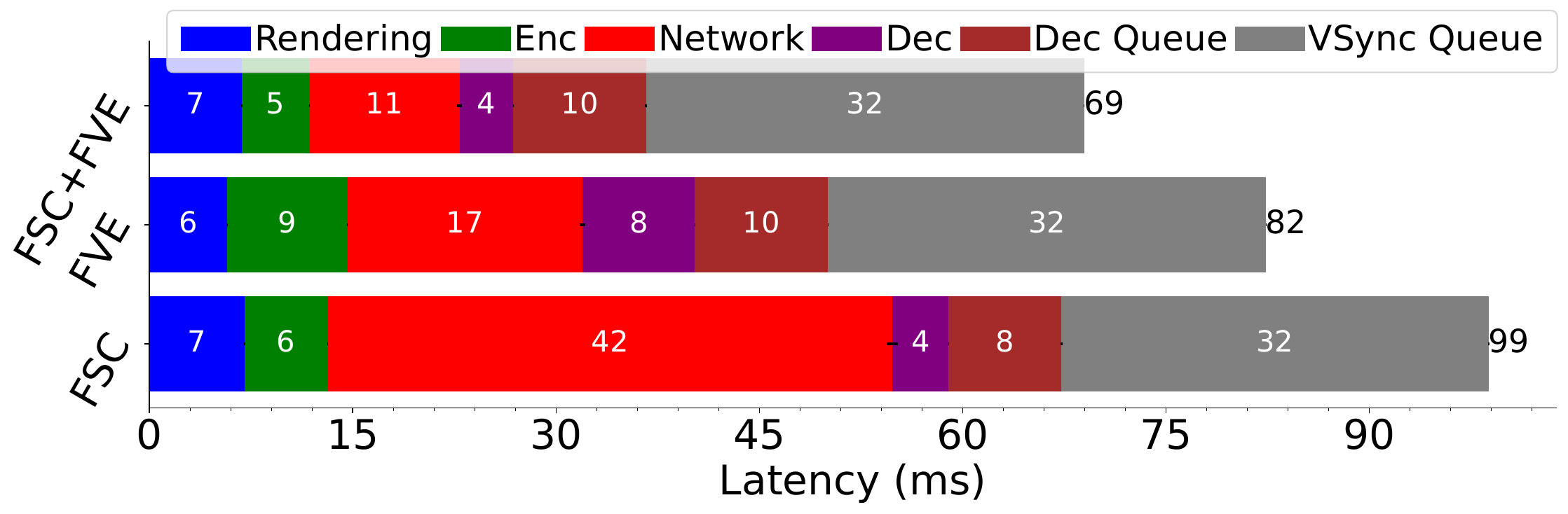}
        \caption{End-to-end Latency Decomposition.}
        \label{fig:ablation_latency}
    \end{subfigure}
    \hfill
    \begin{subfigure}[t]{0.49\textwidth}
        \centering
        \includegraphics[width=.7\linewidth]{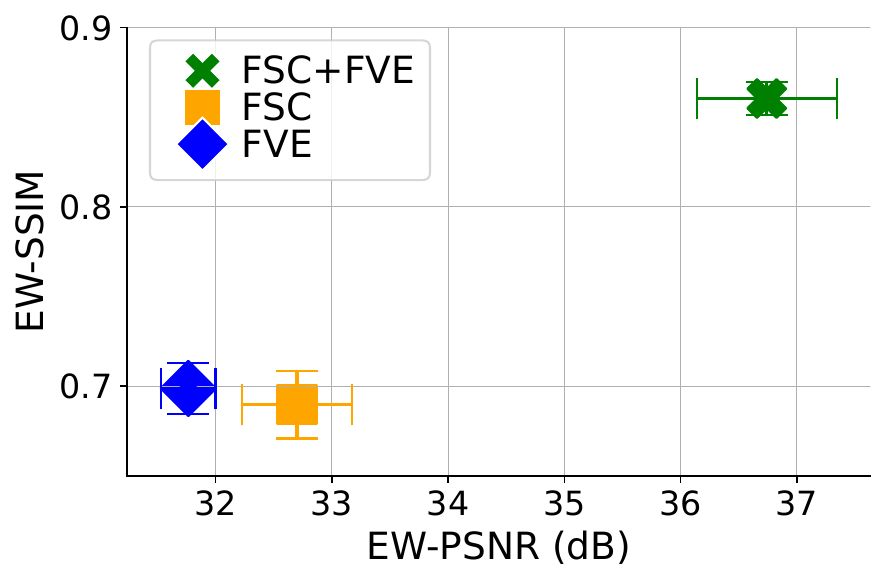}
        \caption{Perceptual Visual Quality (VQ).  }
        \label{fig:ablation_ewssim_ewpsnr}
    \end{subfigure}
    % \vspace{-.2cm}
    \caption{Ablation results. Latency and perceptual visual quality over mobile network. Comparison of FSC (without FVE), and FVE (without FSC) against \sysname (FSC+FVE). (a) FSC increases network latency, while FVE  increases encoding and decoding latency. (b) Decline in VQ when FSC and FVE are used separately.}
    \label{fig:ablation_study}
\end{figure}

We conduct further experiments to assess the individual contributions of foveated spatial compression (FSC) and foveated video encoding (FVE) to an immersive VR gaming experience. We test two setups: (1) \textbf{FVE Only}: \sysname without FSC and (2) \textbf{ FSC Only}: \sysname without FVE (utilizing traditional video encoding). We then compare their performance to \sysname (\textbf{FSC+FVE}), which includes both FSC and FVE over the testbed setup described in~\autoref{sec:setup}.

\textbf{FVE Only.}
As illustrated in~\autoref{fig:ablation_study}, the FVE Only setup shows a significant decline in visual quality, with EWSSIM of 0.70 and EWPSNR of 31.8 dB. In contrast, \sysname (FSC+FVE) achieves an EWSSIM of 0.87 and an EWPSNR of 36.8 dB. The FVE Only configuration increases latency from 69 ms for \sysname (FSC+FVE) to 82 ms. In particular, excluding FSC results in longer video coding times, raising the average encoding latency (Enc) from 5 ms to 9 ms, and average decoding latency (Dec) from 4 ms to 8 ms. This increase is due to the absence of FSC, which leads to higher pixel density and requires processing more macroblocks during video coding. Furthermore, the higher pixel density results in larger encoded frame sizes and higher bitrates, potentially increasing network congestion. Consequently, the FVE-only setup faces a longer network latency (17 ms) compared to \sysname (FSC+FVE) of 11 ms.

\textbf{FSC Only.}
Replacing FVE with traditional video encoding results in uniform quality distribution across frame macroblocks. As illustrated in~\autoref{fig:ablation_study}, this setup increases network latency from 11 ms in \sysname (FSC+FVE) to 42 ms. Additionally, it causes a significant decline in visual quality, with an EWSSIM of 0.68 and an EWPSNR of 32.6 dB, compared to \sysname (FSC+FVE), which has an EWSSIM of 0.87 and an EWPSNR of 36.8 dB. 

\textbf{Findings.} FVE assigns quality in compatibility with HVS, with higher quality to the users’ gaze regions and less in peripheral regions. FVE reduces encoded frame sizes, thus optimizing bandwidth usage and ensuring seamless video streaming. FSC compresses the game frame spatially by assigning a lower pixel density in the periphery. Consequently, FSC accelerates video coding and enables the processing of high-resolution frames using hardware-accelerated codecs. When combined, FVE and FSC enhance perceptual visual quality and player interaction, as demonstrated by the superior performance of \sysname (FSC+FVE) over individual FVE and FSC implementations.

% GCC and CGFVE do not incorporate foveated rendering (FR), which limits their ability to encode game frames at resolutions exceeding 4K (4096 X 4096) according to NVidia\footnote{\url{https://developer.nvidia.com/video-codec-sdk}}. As a result, only ALVR, FovOptix, and \sysname can successfully stream game scenes at resolutions above 4K. While \sysname demonstrates latency comparable to ALVR and FovOptix over WiFi, only \sysname and FovOptix maintain low and acceptable latency under mobile network conditions. Notably, \sysname excels in delivering superior perceived visual quality across all network conditions, surpassing FovOptix.

% \begin{figure}[!t]
%     \centering
%     \begin{subfigure}[b]{.2357\textwidth}
%         \includegraphics[width=1.05\linewidth]{figures/high_res_latency_cdf.png}
%         \caption{CDF of MTP Latency}
%     \label{fig:high_res_cdf}
%     \end{subfigure}
%     \hfill
%      \begin{subfigure}[b]{.2357\textwidth}
%         \includegraphics[width=1.05\linewidth]{figures/high_res_vq.png}
%         \caption{Visual Quality}
%     \label{fig:high_res_vq}
%     \end{subfigure}
    
%     \caption{CDF of MTP Latency and Visual Quality (EWSSIM and EWPSNR) of game frames in higher resolutions over WiFi and Mobile Network setups}
%     \label{fig:high_res_peformance_cdf_vq}
% \end{figure}

\section{User Study}
\label{sec:user}

\begin{figure}[!t]
    \centering
    \begin{subfigure}[b]{.49\textwidth}
        \includegraphics[width=\linewidth]{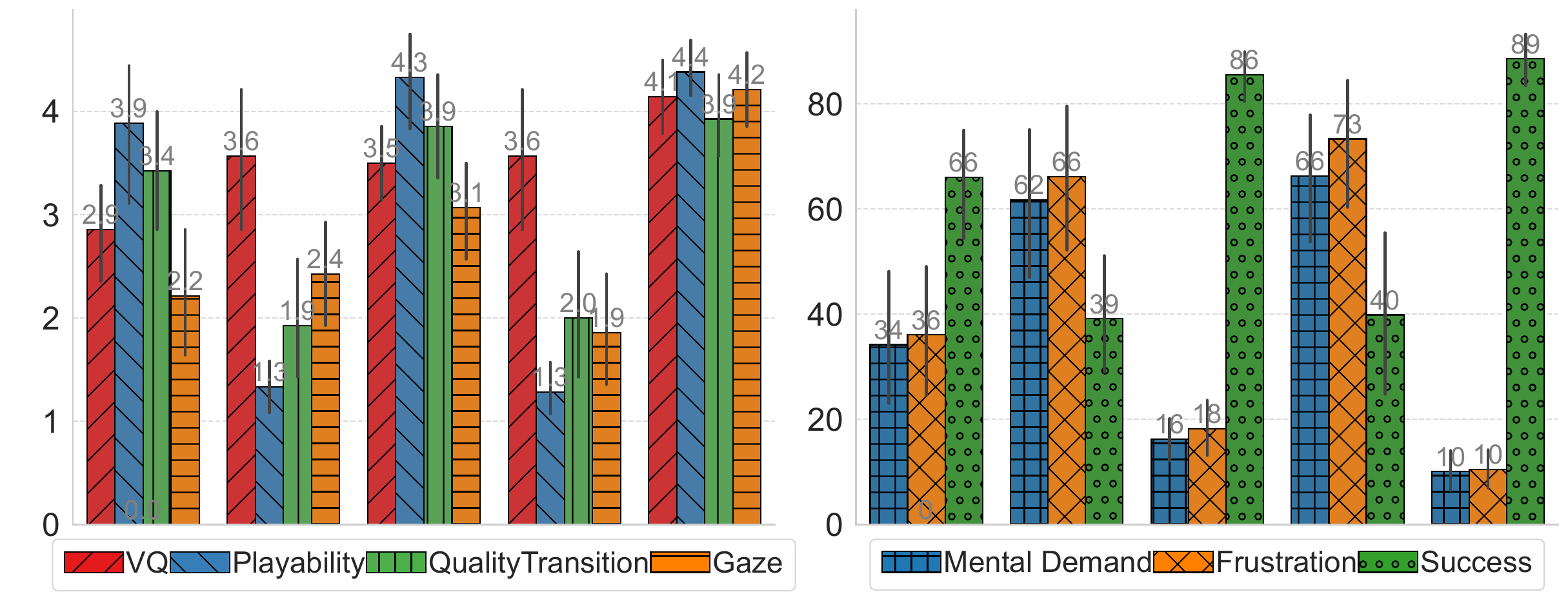}
        \caption{Perception of FPS game (Longbow) }
    \label{fig:mobile_QoE_FPS}
    \end{subfigure}
        \begin{subfigure}[b]{.49\textwidth}
        \includegraphics[width=\linewidth]{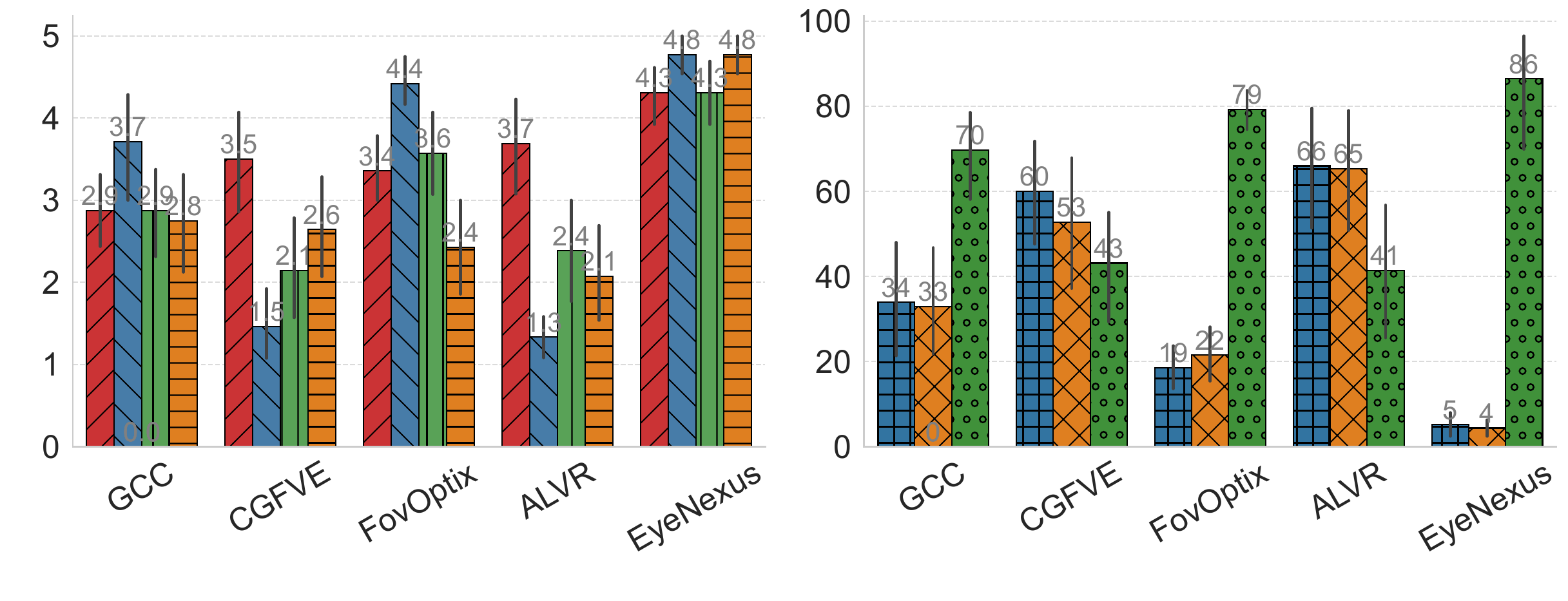}
        \caption{Perception of TPS game (Xortex)}
        \label{fig:mobile_QoE_TPS}
    \end{subfigure}
    % \vspace{-.2cm}
    \caption{Perception of gaming experience under typical measures (left) and task load (right) over 5G network }
     \label{fig:mobile_QoE}
\end{figure}

We conduct an IRB-approved user study to evaluate the effectiveness from the users’ perspective.

%Participants, Apparatus, and Framework
\noindent\textbf{Study Design:} 
We recruit 14 participants (Female: 6, Male: 8), with an average age of $24.4\pm2$. We ask the participants to wear Meta Quest Pro and play two VR Games: Longbow (FPS game) and Xortex (TPS game). Heatmap of participants’ gaze points for both games is shown in~\autoref{fig:heatmap_comparison}. Each game was streamed using the benchmarks defined in~\autoref{tab:baselines} over an emulation of a 5G mobile network. The participants took inter-session breaks to avoid cumulative effects. We restarted the network emulation script (\autoref{sec:testbed}) for each benchmark to maintain consistent network conditions across the benchmarks. Besides, we anonymized the benchmarks and randomized their sequence to prevent learning, order effects, and rating bias. After each gameplay session, %the participant rated their gaming and workload experiences, and susceptibility to motion sickness. Specifically, they 
the participants filled out a questionnaire on a 5-point Likert scale on the perceived visual quality (VQ) and quality gradients (QualityTransition), suitability to visual experience (Gaze), and playability.  They also filled out a NASA TLX ~\cite{hart2006nasa} survey on the perceived mental demand, frustration, and success on a [0-100] scale, and reported susceptibility to motion sickness on a [1-7] scale.% \textit{(1:no symptoms to 7:experiencing vomiting)}. 

% \textit{(1:bad to 5:excellent)}, suitability to visual experience (Gaze) \textit{( 1:not well to 5:Extremely well)}, and playability \textit{(1:laggy to 5:extremely responsive)}.  They also filled out a NASA TLX ~\cite{hart2006nasa} survey on the perceived mental demand, frustration, and success on a [0-100] scale. Besides, they reported their susceptibility to motion sickness on a [1-7] scale \textit{(1:no symptoms to 7:experiencing vomiting)}.

%Results
\noindent\textbf{Results:} 
Figure~\ref{fig:mobile_QoE} shows results with CI 95\% as error bars. Following, we study the results from three perspectives: gaming experience, workload experience, and experience of motion sickness.

%Subsection: Gaming Experience
\subsection{Gaming Experience}
\sysname shows superior performance in both FPS and TPS game genres, achieving the highest perceived visual quality (VQ), playability, and gaze suitability, and providing natural and unnoticeable quality transition. 
%Performance in FPS game
In the FPS game (\autoref{fig:mobile_QoE_FPS}), it outperforms benchmarks with averages of 4.1 for VQ, 4.4 for playability, 3.9 for Quality Transition, and 4.2 for gaze suitability. FovOptix follows with averages of 4.3 for playability, 3.9 for Quality Transition, and 3.1 for gaze suitability. ALVR and CGFVE slightly surpass FovOptix in VQ (mean 3.6 each) but score lower in other metrics. Their low playability ratings stem from rate control issues: CGFVE lacks adaptive rate control, simply reducing sending bitrate, while ALVR's basic rate control inadequately responds to mobile network fluctuations. GCC ranks next in playability (mean 3.9) because its real-time communication design prioritizes low latency over high VQ, leading to the lowest VQ and second lowest gaze compatibility. It does achieve the third-highest score in Quality Transition due to its uniformity.

%Performance in TPS game
In the TPS game (\autoref{fig:mobile_QoE_TPS}), \sysname outperforms all baselines, achieving averages of 4.3 for Visual Quality, 4.8 for Playability, 4.2 for Quality Transition, and 4.8 for Gaze Suitability. FovOptix follows with Playability and Quality Transition averages of 4.4 and 3.6, respectively, but its gaze compatibility (mean 2.4) is lower than GCC (mean 2.8) and CGFVE (mean 2.6). FovOptix's Visual Quality (mean 3.4) also falls short compared to ALVR (mean 3.7) and CGFVE (mean 3.5), while GCC has the lowest Visual Quality at 2.9. FovOptix's performance declines in all aspects except playability, as its fixed foveation is ineffective for frequent gaze shifts from the central FoV.

%Reasoning and Findings of EyeNexus Superiority
One-way ANOVA tests show significant differences ($p<0.005$) among the benchmarks. \sysname's superior performance is due to its real-time eye gaze-based quality assignment, which maintains quality in the fovea and reduces it in the periphery to adapt to network bandwidth, enhancing playability. The Gaussian distribution of quality ensures smooth transitions.

%Subsection: Workload Experience
\subsection{Workload Experience}
\sysname shows superior performance in both FPS and TPS game genres, achieving the lowest frustration and highest perceived success, and requiring the lowest mental efforts. 

%Performance in FPS game
In the FPS game (\autoref{fig:mobile_QoE_FPS}), \sysname outperforms benchmarks with the lowest frustration and mental demand averages (mean 10 each) and the highest perceived success (mean 89). FovOptix follows with averages of 16 for mental demand, 18 for frustration, and 86 for perceived success, while GCC has averages of 34, 36, and 66, respectively. CGFVE and ALVR show the poorest performance, with averages of 62 and 66 for mental demand, 66 and 73 for frustration, and 39 and 40 for perceived success, respectively. 
%Performance in TPS game
In the TPS game (\autoref{fig:mobile_QoE_TPS}), \sysname outperforms benchmarks with the lowest frustration and mental demand averages (mean 5, 4) and the highest perceived success (mean 86). FovOptix follows with averages of 19 for mental demand, 22 for frustration, and 79 for perceived success, while GCC has averages of 34, 33, and 70, respectively. CGFVE and ALVR show the poorest performance, with averages of 60 and 66 for mental demand, 53 and 65 for frustration, and 43 and 41 for perceived success, respectively. 

%In terms of motion sickness, 
The participants achieved the highest game score using \sysname, followed by FovOptix. This aligns with perceived playability, where \sysname also ranks highest. Additionally, \sysname has the lowest motion sickness severity (mean 1), indicating no susceptibility. FovOptix follows for both FPS and TPS games (means of 1.14 and 1.21), while GCC shows slightly higher scores (means of 1.28 and 1.31), indicating minimal discomfort. In contrast, participants reported mild dizziness with ALVR and CGFVE, with means of 2 and 2.21 for FPS games and 2.07 and 2.43 for TPS games.%, respectively.

\section{Findings and Discussion}
\label{sec:discus}
%\sysname provides gaze-contingent spatial compression and video streaming for high QoE over variable network conditions. 
We evaluated \sysname using FPS and TPS games, over WiFi and 5G mobile networks, to understand its performance for predominant game genres over real-life network setups.   

% \textbf{Performance across Various Configurations.} 
\sysname exhibits the lowest latency and the highest perceptual visual quality across major game genres and real-world network setups. \sysname scales up with increasing resolution and network throughput. It shows a linear latency increase and marginal visual quality decrease with a significant resolution increase over variable unstable network conditions (mobile network). It also shows a linear perceptual visual quality increase with the resolution increase. This underscores its potential to meet the growing demand for higher resolution, rooted in \sysname's capability to effectively adapt to network fluctuations while ensuring enhanced visual quality.

% \textbf{Comparison and Reasoning.}
FovOptix performs slightly lower than \sysname in FPS games but offers inferior visual quality and user experience in TPS games. This is attributed to \sysname’s gaze-contingent design, which leverages real-time gaze data to optimize the perceptual visual quality. ALVR and CGFVE utilize partial foveation, which fail to maintain satisfactory QoE on mobile networks. While GCC achieves competitive MTP latency on both Wi-Fi and mobile networks, it experiences a significant drop in visual quality on mobile due to its sensitivity to latency variations and packet loss. To address these fluctuations, GCC reduces the sending bitrate to maintain real-time communication. In contrast, \sysname excels in delivering improved perceived visual quality, a smooth quality gradient, and seamless interaction, making it the ideal choice for optimal QoE over variable network conditions.

\section{Conclusion}
\label{sec:conclusion}
This paper presented \sysname, a gaze-contingent spatial compression (FSC) and video streaming (FVE) system that leverages the non-uniformity in human vision.  \sysname features a novel foveation controller that adjusts quality according to a Gaussian distribution based on real-time gaze data and bandwidth usage. We evaluated \sysname over major network setups and game genres. Our findings revealed that \sysname reduces latency and enhances visual quality (VQ) compared to the Google standard and research SoTA. The latency and VQ scale up for increased resolutions and over various network settings. Our IRB-approved user study revealed that \sysname achieves the highest playability and perceived VQ while eliminating the feeling of motion sickness.
 
\begin{acks}
This research was supported in part by a grant from the Guangzhou Municipal Nansha District Science and Technology Bureau under Contract No.2022ZD01 and the MetaHKUST project from the Hong Kong University of Science and Technology (Guangzhou).
\end{acks}

\bibliographystyle{ACM-Reference-Format}
\bibliography{references}


\begin{thebibliography}{45}

%%% the Web address.

\ifx \showCODEN    \undefined \def \showCODEN     #1{\unskip}     \fi
\ifx \showDOI      \undefined \def \showDOI       #1{#1}\fi
\ifx \showISBNx    \undefined \def \showISBNx     #1{\unskip}     \fi
\ifx \showISBNxiii \undefined \def \showISBNxiii  #1{\unskip}     \fi
\ifx \showISSN     \undefined \def \showISSN      #1{\unskip}     \fi
\ifx \showLCCN     \undefined \def \showLCCN      #1{\unskip}     \fi
\ifx \shownote     \undefined \def \shownote      #1{#1}          \fi
\ifx \showarticletitle \undefined \def \showarticletitle #1{#1}   \fi
\ifx \showURL      \undefined \def \showURL       {\relax}        \fi
% invisible to TeX
\providecommand\bibfield[2]{#2}
\providecommand\bibinfo[2]{#2}
\providecommand\natexlab[1]{#1}
\providecommand\showeprint[2][]{arXiv:#2}

\bibitem[web(2023)]%
        {webrtc}
 \bibinfo{year}{2023}\natexlab{}.
\newblock \bibinfo{title}{{Real-time Communication for the Web (WebRTC)}}.
\newblock
\newblock
\urldef\tempurl%
\url{https://webrtc.org/}
\showURL{%
\tempurl}


\bibitem[alv(2024)]%
        {alvr}
 \bibinfo{year}{2024}\natexlab{}.
\newblock \bibinfo{title}{{Air Light VR (ALVR)}}.
\newblock
\newblock
\urldef\tempurl%
\url{https://github.com/alvr-org/ALVR}
\showURL{%
\tempurl}


\bibitem[ste(2024)]%
        {steamVR}
 \bibinfo{year}{2024}\natexlab{}.
\newblock \bibinfo{title}{{SteamVR}}.
\newblock
\newblock
\urldef\tempurl%
\url{https://store.steampowered.com/app/250820/SteamVR/}
\showURL{%
\tempurl}
\newblock
\shownote{Access date: October 25, 2023}.


\bibitem[VRg(2024)]%
        {VRgamingMarket2024}
 \bibinfo{year}{2024}\natexlab{}.
\newblock \bibinfo{title}{Virtual Reality (VR) in Gaming Market Size, Share \& Industry Analysis}.
\newblock
\newblock
\urldef\tempurl%
\url{https://www.fortunebusinessinsights.com/industry-reports/virtual-reality-gaming-market-100271}
\showURL{%
\tempurl}


\bibitem[Alhilal et~al\mbox{.}(2022)]%
        {alhilal2022nebula}
\bibfield{author}{\bibinfo{person}{Ahmad Alhilal}, \bibinfo{person}{Tristan Braud}, \bibinfo{person}{Bo Han}, {and} \bibinfo{person}{Pan Hui}.} \bibinfo{year}{2022}\natexlab{}.
\newblock \showarticletitle{Nebula: Reliable Low-Latency Video Transmission for Mobile Cloud Gaming}. In \bibinfo{booktitle}{\emph{Proceedings of the ACM Web Conference 2022}} (Virtual Event, Lyon, France) \emph{(\bibinfo{series}{WWW '22})}. \bibinfo{publisher}{Association for Computing Machinery}, \bibinfo{address}{New York, NY, USA}, \bibinfo{pages}{3407–3417}.
\newblock
\showISBNx{9781450390965}
\urldef\tempurl%
\url{https://doi.org/10.1145/3485447.3512276}
\showDOI{\tempurl}


\bibitem[Alhilal et~al\mbox{.}(2024)]%
        {alhilal2024FovOptix}
\bibfield{author}{\bibinfo{person}{Ahmad Alhilal}, \bibinfo{person}{Ze Wu}, \bibinfo{person}{Yuk~Hang Tsui}, {and} \bibinfo{person}{Pan Hui}.} \bibinfo{year}{2024}\natexlab{}.
\newblock \showarticletitle{FovOptix: Human Vision-Compatible Video Encoding and Adaptive Streaming in VR Cloud Gaming}. In \bibinfo{booktitle}{\emph{Proceedings of the 15th ACM Multimedia Systems Conference}} (Bari, Italy) \emph{(\bibinfo{series}{MMSys '24})}. \bibinfo{publisher}{Association for Computing Machinery}, \bibinfo{address}{New York, NY, USA}, \bibinfo{pages}{67–77}.
\newblock
\showISBNx{9798400704123}
\urldef\tempurl%
\url{https://doi.org/10.1145/3625468.3647612}
\showDOI{\tempurl}


\bibitem[{Android Developers}(2023)]%
        {android_media_codec}
\bibfield{author}{\bibinfo{person}{{Android Developers}}.} \bibinfo{year}{2023}\natexlab{}.
\newblock \bibinfo{title}{MediaCodec | Android Developers}.
\newblock
\newblock
\urldef\tempurl%
\url{https://developer.android.com/reference/android/media/MediaCodec}
\showURL{%
\tempurl}
\newblock
\shownote{Accessed: 2023-12-03}.


\bibitem[Bolot and Turletti(1994)]%
        {feedbackpkt}
\bibfield{author}{\bibinfo{person}{J.-C. Bolot} {and} \bibinfo{person}{T. Turletti}.} \bibinfo{year}{1994}\natexlab{}.
\newblock \showarticletitle{A rate control mechanism for packet video in the Internet}. In \bibinfo{booktitle}{\emph{Proceedings of INFOCOM '94 Conference on Computer Communications}}. \bibinfo{pages}{1216--1223 vol.3}.
\newblock
\urldef\tempurl%
\url{https://doi.org/10.1109/INFCOM.1994.337568}
\showDOI{\tempurl}


\bibitem[Carlucci et~al\mbox{.}(2016a)]%
        {GCCwebrtc}
\bibfield{author}{\bibinfo{person}{Gaetano Carlucci}, \bibinfo{person}{Luca De~Cicco}, \bibinfo{person}{Stefan Holmer}, {and} \bibinfo{person}{Saverio Mascolo}.} \bibinfo{year}{2016}\natexlab{a}.
\newblock \showarticletitle{Analysis and Design of the Google Congestion Control for Web Real-Time Communication (WebRTC)}. In \bibinfo{booktitle}{\emph{Proceedings of the 7th International Conference on Multimedia Systems}} (Klagenfurt, Austria) \emph{(\bibinfo{series}{MMSys '16})}. \bibinfo{publisher}{Association for Computing Machinery}, \bibinfo{address}{New York, NY, USA}, Article \bibinfo{articleno}{13}, \bibinfo{numpages}{12}~pages.
\newblock
\showISBNx{9781450342971}
\urldef\tempurl%
\url{https://doi.org/10.1145/2910017.2910605}
\showDOI{\tempurl}


\bibitem[Carlucci et~al\mbox{.}(2017)]%
        {queuingdelaygradient2}
\bibfield{author}{\bibinfo{person}{Gaetano Carlucci}, \bibinfo{person}{Luca De~Cicco}, \bibinfo{person}{Stefan Holmer}, {and} \bibinfo{person}{Saverio Mascolo}.} \bibinfo{year}{2017}\natexlab{}.
\newblock \showarticletitle{Congestion Control for Web Real-Time Communication}.
\newblock \bibinfo{journal}{\emph{IEEE/ACM Transactions on Networking}} \bibinfo{volume}{25}, \bibinfo{number}{5} (\bibinfo{year}{2017}), \bibinfo{pages}{2629--2642}.
\newblock
\urldef\tempurl%
\url{https://doi.org/10.1109/TNET.2017.2703615}
\showDOI{\tempurl}


\bibitem[Carlucci et~al\mbox{.}(2016b)]%
        {gccAnalysis}
\bibfield{author}{\bibinfo{person}{Gaetano Carlucci}, \bibinfo{person}{Luca De~Cicco}, \bibinfo{person}{Cesar Ilharco}, {and} \bibinfo{person}{Saverio Mascolo}.} \bibinfo{year}{2016}\natexlab{b}.
\newblock \showarticletitle{Congestion control for real-time communications: A comparison between NADA and GCC}. In \bibinfo{booktitle}{\emph{2016 24th Mediterranean Conference on Control and Automation (MED)}}. \bibinfo{pages}{575--580}.
\newblock
\urldef\tempurl%
\url{https://doi.org/10.1109/MED.2016.7535978}
\showDOI{\tempurl}


\bibitem[Corporation(2023)]%
        {nvidia_nvenc_2023}
\bibfield{author}{\bibinfo{person}{NVIDIA Corporation}.} \bibinfo{year}{2023}\natexlab{}.
\newblock \bibinfo{title}{NVENC Video Encoder API Programming Guide}.
\newblock
\newblock
\urldef\tempurl%
\url{https://docs.nvidia.com/video-technologies/video-codec-sdk/12.1/nvenc-video-encoder-api-prog-guide/index.html}
\showURL{%
\tempurl}
\newblock
\shownote{Accessed: 2024-11-30}.


\bibitem[Cui et~al\mbox{.}(2023)]%
        {feedbackpkt1}
\bibfield{author}{\bibinfo{person}{Changjiang Cui}, \bibinfo{person}{Yifei Lu}, \bibinfo{person}{Zhen Wang}, \bibinfo{person}{Zeqi Ruan}, {and} \bibinfo{person}{Hongxiang Wang}.} \bibinfo{year}{2023}\natexlab{}.
\newblock \showarticletitle{MM-ABR: an Enhanced ABR Algorithm with Multi-Metric Information for QUIC-based Video Streaming}. In \bibinfo{booktitle}{\emph{2023 IEEE 29th International Conference on Parallel and Distributed Systems (ICPADS)}}. \bibinfo{pages}{1214--1221}.
\newblock
\urldef\tempurl%
\url{https://doi.org/10.1109/ICPADS60453.2023.00176}
\showDOI{\tempurl}


\bibitem[Fang et~al\mbox{.}(2023)]%
        {fang23nossdav}
\bibfield{author}{\bibinfo{person}{Jia-Wei Fang}, \bibinfo{person}{Kuan-Yu Lee}, \bibinfo{person}{Teemu K\"{a}m\"{a}r\"{a}inen}, \bibinfo{person}{Matti Siekkinen}, {and} \bibinfo{person}{Cheng-Hsin Hsu}.} \bibinfo{year}{2023}\natexlab{}.
\newblock \showarticletitle{Will Dynamic Foveation Boost Cloud VR Gaming Experience?}. In \bibinfo{booktitle}{\emph{Proceedings of the 33rd Workshop on Network and Operating System Support for Digital Audio and Video}} (Vancouver, BC, Canada) \emph{(\bibinfo{series}{NOSSDAV '23})}. \bibinfo{publisher}{Association for Computing Machinery}, \bibinfo{address}{New York, NY, USA}, \bibinfo{pages}{29–35}.
\newblock
\showISBNx{9798400701849}
\urldef\tempurl%
\url{https://doi.org/10.1145/3592473.3592565}
\showDOI{\tempurl}


\bibitem[FFmpeg(2023)]%
        {FFmpegH264}
\bibfield{author}{\bibinfo{person}{FFmpeg}.} \bibinfo{year}{2023}\natexlab{}.
\newblock \bibinfo{title}{Encode/H.264}.
\newblock
\newblock
\urldef\tempurl%
\url{https://trac.ffmpeg.org/wiki/Encode/H.264}
\showURL{%
\tempurl}
\newblock
\shownote{Accessed: 2024-12-04}.


\bibitem[Frieß et~al\mbox{.}(2021)]%
        {foveatedEncodingForLargeResolutionDisplays}
\bibfield{author}{\bibinfo{person}{Florian Frieß}, \bibinfo{person}{Matthias Braun}, \bibinfo{person}{Valentin Bruder}, \bibinfo{person}{Steffen Frey}, \bibinfo{person}{Guido Reina}, {and} \bibinfo{person}{Thomas Ertl}.} \bibinfo{year}{2021}\natexlab{}.
\newblock \showarticletitle{Foveated Encoding for Large High-Resolution Displays}.
\newblock \bibinfo{journal}{\emph{IEEE Transactions on Visualization and Computer Graphics}} \bibinfo{volume}{27}, \bibinfo{number}{2} (\bibinfo{year}{2021}), \bibinfo{pages}{1850--1859}.
\newblock
\urldef\tempurl%
\url{https://doi.org/10.1109/TVCG.2020.3030445}
\showDOI{\tempurl}


\bibitem[Group(2024a)]%
        {OpenGL}
\bibfield{author}{\bibinfo{person}{Khronos Group}.} \bibinfo{year}{2024}\natexlab{a}.
\newblock \bibinfo{title}{OpenGL - The Industry Standard for High Performance Graphics}.
\newblock
\newblock
\urldef\tempurl%
\url{https://www.opengl.org/}
\showURL{%
\tempurl}
\newblock
\shownote{Accessed: 2024-12-03}.


\bibitem[Group(2024b)]%
        {OpenXR}
\bibfield{author}{\bibinfo{person}{Khronos Group}.} \bibinfo{year}{2024}\natexlab{b}.
\newblock \bibinfo{title}{OpenXR Specification}.
\newblock
\newblock
\urldef\tempurl%
\url{https://registry.khronos.org/OpenXR/specs/1.0/html/xrspec.html#_what_is_openxr}
\showURL{%
\tempurl}
\newblock
\shownote{Version 1.0.34}.


\bibitem[Hart(2006)]%
        {hart2006nasa}
\bibfield{author}{\bibinfo{person}{Sandra~G Hart}.} \bibinfo{year}{2006}\natexlab{}.
\newblock \showarticletitle{NASA-task load index (NASA-TLX); 20 years later}. In \bibinfo{booktitle}{\emph{Proceedings of the human factors and ergonomics society annual meeting}}, Vol.~\bibinfo{volume}{50}. Sage publications Sage CA: Los Angeles, CA, \bibinfo{pages}{904--908}.
\newblock


\bibitem[Hayes and Armitage(2011)]%
        {queuingdelaygradient4}
\bibfield{author}{\bibinfo{person}{David~A. Hayes} {and} \bibinfo{person}{Grenville Armitage}.} \bibinfo{year}{2011}\natexlab{}.
\newblock \showarticletitle{Revisiting TCP congestion control using delay gradients}. In \bibinfo{booktitle}{\emph{Proceedings of the 10th International IFIP TC 6 Conference on Networking - Volume Part II}} (Valencia, Spain) \emph{(\bibinfo{series}{NETWORKING'11})}. \bibinfo{publisher}{Springer-Verlag}, \bibinfo{address}{Berlin, Heidelberg}, \bibinfo{pages}{328–341}.
\newblock
\showISBNx{9783642207976}


\bibitem[Huang et~al\mbox{.}(2018)]%
        {queuingdelaygradient1}
\bibfield{author}{\bibinfo{person}{Tianchi Huang}, \bibinfo{person}{Rui-Xiao Zhang}, \bibinfo{person}{Chao Zhou}, {and} \bibinfo{person}{Lifeng Sun}.} \bibinfo{year}{2018}\natexlab{}.
\newblock \showarticletitle{QARC: Video Quality Aware Rate Control for Real-Time Video Streaming based on Deep Reinforcement Learning}. In \bibinfo{booktitle}{\emph{Proceedings of the 26th ACM International Conference on Multimedia}} (Seoul, Republic of Korea) \emph{(\bibinfo{series}{MM '18})}. \bibinfo{publisher}{Association for Computing Machinery}, \bibinfo{address}{New York, NY, USA}, \bibinfo{pages}{1208–1216}.
\newblock
\showISBNx{9781450356657}
\urldef\tempurl%
\url{https://doi.org/10.1145/3240508.3240545}
\showDOI{\tempurl}


\bibitem[Illahi et~al\mbox{.}(2020)]%
        {Illahi2020CGFVE}
\bibfield{author}{\bibinfo{person}{Gazi~Karam Illahi}, \bibinfo{person}{Thomas~Van Gemert}, \bibinfo{person}{Matti Siekkinen}, \bibinfo{person}{Enrico Masala}, \bibinfo{person}{Antti Oulasvirta}, {and} \bibinfo{person}{Antti Yl\"{a}-J\"{a}\"{a}ski}.} \bibinfo{year}{2020}\natexlab{}.
\newblock \showarticletitle{Cloud Gaming with Foveated Video Encoding}.
\newblock  \bibinfo{volume}{16}, \bibinfo{number}{1}, Article \bibinfo{articleno}{7} (\bibinfo{date}{Feb.} \bibinfo{year}{2020}), \bibinfo{numpages}{24}~pages.
\newblock
\showISSN{1551-6857}
\urldef\tempurl%
\url{https://doi.org/10.1145/3369110}
\showDOI{\tempurl}


\bibitem[Illahi et~al\mbox{.}(2021)]%
        {illahi21mmsys}
\bibfield{author}{\bibinfo{person}{Gazi~Karam Illahi}, \bibinfo{person}{Matti Siekkinen}, \bibinfo{person}{Teemu K\"{a}m\"{a}r\"{a}inen}, {and} \bibinfo{person}{Antti Yl\"{a}-J\"{a}\"{a}ski}.} \bibinfo{year}{2021}\natexlab{}.
\newblock \showarticletitle{Foveated streaming of real-time graphics}. In \bibinfo{booktitle}{\emph{Proceedings of the 12th ACM Multimedia Systems Conference}} (Istanbul, Turkey) \emph{(\bibinfo{series}{MMSys '21})}. \bibinfo{publisher}{Association for Computing Machinery}, \bibinfo{address}{New York, NY, USA}, \bibinfo{pages}{214–226}.
\newblock
\showISBNx{9781450384346}
\urldef\tempurl%
\url{https://doi.org/10.1145/3458305.3463383}
\showDOI{\tempurl}


\bibitem[K\"{a}m\"{a}r\"{a}inen and Siekkinen(2023a)]%
        {kamarainen23metasys}
\bibfield{author}{\bibinfo{person}{Teemu K\"{a}m\"{a}r\"{a}inen} {and} \bibinfo{person}{Matti Siekkinen}.} \bibinfo{year}{2023}\natexlab{a}.
\newblock \showarticletitle{Foveated Spatial Compression for Remote Rendered Virtual Reality}. In \bibinfo{booktitle}{\emph{Proceedings of the First Workshop on Metaverse Systems and Applications}} (Helsinki, Finland) \emph{(\bibinfo{series}{MetaSys '23})}. \bibinfo{publisher}{Association for Computing Machinery}, \bibinfo{address}{New York, NY, USA}, \bibinfo{pages}{7–13}.
\newblock
\showISBNx{9798400702136}
\urldef\tempurl%
\url{https://doi.org/10.1145/3597063.3597359}
\showDOI{\tempurl}


\bibitem[K\"{a}m\"{a}r\"{a}inen and Siekkinen(2023b)]%
        {foveatedspatialcompression}
\bibfield{author}{\bibinfo{person}{Teemu K\"{a}m\"{a}r\"{a}inen} {and} \bibinfo{person}{Matti Siekkinen}.} \bibinfo{year}{2023}\natexlab{b}.
\newblock \showarticletitle{Foveated Spatial Compression for Remote Rendered Virtual Reality}. In \bibinfo{booktitle}{\emph{Proceedings of the First Workshop on Metaverse Systems and Applications}} (Helsinki, Finland) \emph{(\bibinfo{series}{MetaSys '23})}. \bibinfo{publisher}{Association for Computing Machinery}, \bibinfo{address}{New York, NY, USA}, \bibinfo{pages}{7–13}.
\newblock
\showISBNx{9798400702136}
\urldef\tempurl%
\url{https://doi.org/10.1145/3597063.3597359}
\showDOI{\tempurl}


\bibitem[Li et~al\mbox{.}(2011)]%
        {li2011visual}
\bibfield{author}{\bibinfo{person}{Zhicheng Li}, \bibinfo{person}{Shiyin Qin}, {and} \bibinfo{person}{Laurent Itti}.} \bibinfo{year}{2011}\natexlab{}.
\newblock \showarticletitle{Visual attention guided bit allocation in video compression}.
\newblock \bibinfo{journal}{\emph{Image and Vision Computing}} \bibinfo{volume}{29}, \bibinfo{number}{1} (\bibinfo{year}{2011}), \bibinfo{pages}{1--14}.
\newblock


\bibitem[Meta(2019)]%
        {AADTOculus}
\bibfield{author}{\bibinfo{person}{Meta}.} \bibinfo{year}{2019}\natexlab{}.
\newblock \bibinfo{title}{How does Oculus Link Work? The Architecture, Pipeline and AADT Explained}.
\newblock
\newblock
\urldef\tempurl%
\url{https://developer.oculus.com/blog/howdoes-oculus-link-work-the-architecture-pipeline-and-aadt-explained/}
\showURL{%
\tempurl}
\newblock
\shownote{Retrieved December 5, 2024}.


\bibitem[{Meta Horizon OS Developers}(2023)]%
        {meta_horizon_vr_frames}
\bibfield{author}{\bibinfo{person}{{Meta Horizon OS Developers}}.} \bibinfo{year}{2023}\natexlab{}.
\newblock \bibinfo{title}{A VR Frame's Life}.
\newblock
\newblock
\urldef\tempurl%
\url{https://developers.meta.com/horizon/blog/a-vr-frames-life/?hmsr=joyk.com&utm_source=joyk.com&utm_medium=referral&locale=tr_TR}
\showURL{%
\tempurl}
\newblock
\shownote{Accessed: 2023-12-03}.


\bibitem[{NVIDIA}(2023)]%
        {nvidia_video_codec_sdk}
\bibfield{author}{\bibinfo{person}{{NVIDIA}}.} \bibinfo{year}{2023}\natexlab{}.
\newblock \bibinfo{title}{Video Codec SDK}.
\newblock
\newblock
\urldef\tempurl%
\url{https://developer.nvidia.com/video-codec-sdk}
\showURL{%
\tempurl}
\newblock
\shownote{Accessed: 2024-11-30}.


\bibitem[NVIDIA(2024)]%
        {EmphasisMAP_nvenc}
\bibfield{author}{\bibinfo{person}{NVIDIA}.} \bibinfo{year}{2024}\natexlab{}.
\newblock \bibinfo{title}{Emphasis MAP in NVENCODE API}.
\newblock
\newblock
\urldef\tempurl%
\url{https://docs.nvidia.com/video-technologies/video-codec-sdk/11.1/nvenc-video-encoder-api-prog-guide/index.html#emphasis-map}
\showURL{%
\tempurl}


\bibitem[Patney et~al\mbox{.}(2016)]%
        {patney16foveation}
\bibfield{author}{\bibinfo{person}{Anjul Patney}, \bibinfo{person}{Marco Salvi}, \bibinfo{person}{Joohwan Kim}, \bibinfo{person}{Anton Kaplanyan}, \bibinfo{person}{Chris Wyman}, \bibinfo{person}{Nir Benty}, \bibinfo{person}{David Luebke}, {and} \bibinfo{person}{Aaron Lefohn}.} \bibinfo{year}{2016}\natexlab{}.
\newblock \showarticletitle{Towards foveated rendering for gaze-tracked virtual reality}.
\newblock \bibinfo{journal}{\emph{ACM Trans. Graph.}} \bibinfo{volume}{35}, \bibinfo{number}{6}, Article \bibinfo{articleno}{179} (\bibinfo{date}{Dec.} \bibinfo{year}{2016}), \bibinfo{numpages}{12}~pages.
\newblock
\showISSN{0730-0301}
\urldef\tempurl%
\url{https://doi.org/10.1145/2980179.2980246}
\showDOI{\tempurl}


\bibitem[{PixelTools}(2022)]%
        {pixeltools_rate_control}
\bibfield{author}{\bibinfo{person}{{PixelTools}}.} \bibinfo{year}{2022}\natexlab{}.
\newblock \bibinfo{title}{Theoretical discussion on achieving rate control during encoding}.
\newblock
\newblock
\urldef\tempurl%
\url{https://www.pixeltools.com/rate_control_paper.html}
\showURL{%
\tempurl}
\newblock
\shownote{Accessed: 2024-12-02}.


\bibitem[Raca et~al\mbox{.}(2020)]%
        {raca2020beyond}
\bibfield{author}{\bibinfo{person}{Darijo Raca}, \bibinfo{person}{Dylan Leahy}, \bibinfo{person}{Cormac~J. Sreenan}, {and} \bibinfo{person}{Jason~J. Quinlan}.} \bibinfo{year}{2020}\natexlab{}.
\newblock \showarticletitle{Beyond Throughput, the next Generation: A 5G Dataset with Channel and Context Metrics}. In \bibinfo{booktitle}{\emph{Proceedings of the 11th ACM Multimedia Systems Conference}} (Istanbul, Turkey) \emph{(\bibinfo{series}{MMSys '20})}. \bibinfo{publisher}{Association for Computing Machinery}, \bibinfo{address}{New York, NY, USA}, \bibinfo{pages}{303–308}.
\newblock
\showISBNx{9781450368452}
\urldef\tempurl%
\url{https://doi.org/10.1145/3339825.3394938}
\showDOI{\tempurl}


\bibitem[Rimac-Drlje et~al\mbox{.}(2011)]%
        {fassim2011foveation}
\bibfield{author}{\bibinfo{person}{Snježana Rimac-Drlje}, \bibinfo{person}{Goran Martinović}, {and} \bibinfo{person}{Branka Zovko-Cihlar}.} \bibinfo{year}{2011}\natexlab{}.
\newblock \showarticletitle{Foveation-based content Adaptive Structural Similarity index}. In \bibinfo{booktitle}{\emph{2011 18th International Conference on Systems, Signals and Image Processing}}. \bibinfo{pages}{1--4}.
\newblock


\bibitem[Rolff et~al\mbox{.}(2023)]%
        {rolff23vrsnerf}
\bibfield{author}{\bibinfo{person}{Tim Rolff}, \bibinfo{person}{Susanne Schmidt}, \bibinfo{person}{Ke Li}, \bibinfo{person}{Frank Steinicke}, {and} \bibinfo{person}{Simone Frintrop}.} \bibinfo{year}{2023}\natexlab{}.
\newblock \showarticletitle{VRS-NeRF: Accelerating Neural Radiance Field Rendering with Variable Rate Shading}. In \bibinfo{booktitle}{\emph{2023 IEEE International Symposium on Mixed and Augmented Reality (ISMAR)}}. \bibinfo{pages}{243--252}.
\newblock
\urldef\tempurl%
\url{https://doi.org/10.1109/ISMAR59233.2023.00039}
\showDOI{\tempurl}


\bibitem[Silverstein(1996)]%
        {humanvisualsystem}
\bibfield{author}{\bibinfo{person}{Louis~D. Silverstein}.} \bibinfo{year}{1996}\natexlab{}.
\newblock \showarticletitle{Foundations of Vision, by Brian A. Wandell, Sinauer Associates, Inc., Sunderland, MA, 1995. xvi + 476 pp., hardcover \$49.95.}
\newblock \bibinfo{journal}{\emph{Color Research \& Application}} \bibinfo{volume}{21}, \bibinfo{number}{2} (\bibinfo{year}{1996}), \bibinfo{pages}{142--144}.
\newblock
\urldef\tempurl%
\url{https://doi.org/10.1002/col.5080210213}
\showDOI{\tempurl}
\showeprint{https://onlinelibrary.wiley.com/doi/pdf/10.1002/col.5080210213}


\bibitem[Stevewhims({[n.\,d.]})]%
        {Stevewhims}
\bibfield{author}{\bibinfo{person}{Stevewhims}.} \bibinfo{year}{[n.\,d.]}\natexlab{}.
\newblock \bibinfo{title}{Rendering (direct3d 11 graphics) - win32 apps}.
\newblock
\newblock
\urldef\tempurl%
\url{https://learn.microsoft.com/en-us/windows/win32/direct3d11/overviews-direct3d-11-render}
\showURL{%
\tempurl}


\bibitem[Strasburger et~al\mbox{.}(2011)]%
        {degreesforfovea}
\bibfield{author}{\bibinfo{person}{Hans Strasburger}, \bibinfo{person}{Ingo Rentschler}, {and} \bibinfo{person}{Martin Jüttner}.} \bibinfo{year}{2011}\natexlab{}.
\newblock \showarticletitle{Peripheral vision and pattern recognition: A review}.
\newblock \bibinfo{journal}{\emph{Journal of Vision}} \bibinfo{volume}{11}, \bibinfo{number}{5} (\bibinfo{date}{12} \bibinfo{year}{2011}), \bibinfo{pages}{13--13}.
\newblock
\showISSN{1534-7362}
\urldef\tempurl%
\url{https://doi.org/10.1167/11.5.13}
\showDOI{\tempurl}
\showeprint{https://arvojournals.org/arvo/content\_public/journal/jov/933487/i1534-7362-11-5-13\_1713186075.40385.pdf}


\bibitem[Tadahal et~al\mbox{.}(2021)]%
        {ABRSurvey}
\bibfield{author}{\bibinfo{person}{Sujay~S Tadahal}, \bibinfo{person}{Sushma~V Gummadi}, \bibinfo{person}{Kartik Prajapat}, \bibinfo{person}{S~M Meena}, \bibinfo{person}{Uday Kulkarni}, \bibinfo{person}{Sunil~V Gurlahosur}, {and} \bibinfo{person}{Shashidhara Vyakaranal}.} \bibinfo{year}{2021}\natexlab{}.
\newblock \showarticletitle{A Survey On Adaptive Bitrate Algorithms and Their Improvisations}. In \bibinfo{booktitle}{\emph{2021 International Conference on Intelligent Technologies (CONIT)}}. \bibinfo{pages}{1--7}.
\newblock
\urldef\tempurl%
\url{https://doi.org/10.1109/CONIT51480.2021.9498318}
\showDOI{\tempurl}


\bibitem[Walton et~al\mbox{.}(2021)]%
        {walton21metamers}
\bibfield{author}{\bibinfo{person}{David~R. Walton}, \bibinfo{person}{Rafael Kuffner~Dos Anjos}, \bibinfo{person}{Sebastian Friston}, \bibinfo{person}{David Swapp}, \bibinfo{person}{Kaan Ak\c{s}it}, \bibinfo{person}{Anthony Steed}, {and} \bibinfo{person}{Tobias Ritschel}.} \bibinfo{year}{2021}\natexlab{}.
\newblock \showarticletitle{Beyond blur: real-time ventral metamers for foveated rendering}.
\newblock \bibinfo{journal}{\emph{ACM Trans. Graph.}} \bibinfo{volume}{40}, \bibinfo{number}{4}, Article \bibinfo{articleno}{48} (\bibinfo{date}{July} \bibinfo{year}{2021}), \bibinfo{numpages}{14}~pages.
\newblock
\showISSN{0730-0301}
\urldef\tempurl%
\url{https://doi.org/10.1145/3450626.3459943}
\showDOI{\tempurl}


\bibitem[Wu et~al\mbox{.}(2024)]%
        {Theia}
\bibfield{author}{\bibinfo{person}{Nan Wu}, \bibinfo{person}{Kaiyan Liu}, \bibinfo{person}{Ruizhi Cheng}, \bibinfo{person}{Bo Han}, {and} \bibinfo{person}{Puqi Zhou}.} \bibinfo{year}{2024}\natexlab{}.
\newblock \showarticletitle{Theia: Gaze-driven and Perception-aware Volumetric Content Delivery for Mixed Reality Headsets}. In \bibinfo{booktitle}{\emph{Proceedings of the 22nd Annual International Conference on Mobile Systems, Applications and Services}} (Minato-ku, Tokyo, Japan) \emph{(\bibinfo{series}{MOBISYS '24})}. \bibinfo{publisher}{Association for Computing Machinery}, \bibinfo{address}{New York, NY, USA}, \bibinfo{pages}{70–84}.
\newblock
\showISBNx{9798400705816}
\urldef\tempurl%
\url{https://doi.org/10.1145/3643832.3661858}
\showDOI{\tempurl}


\bibitem[Xiong et~al\mbox{.}(2021)]%
        {vrframecomposition}
\bibfield{author}{\bibinfo{person}{Jianghao Xiong}, \bibinfo{person}{En-Lin Hsiang}, \bibinfo{person}{Ziqian He}, \bibinfo{person}{Tao Zhan}, {and} \bibinfo{person}{Shin-Tson Wu}.} \bibinfo{year}{2021}\natexlab{}.
\newblock \showarticletitle{Augmented reality and virtual reality displays: emerging technologies and future perspectives}.
\newblock \bibinfo{journal}{\emph{Light: Science \& Applications}} \bibinfo{volume}{10}, \bibinfo{number}{1} (\bibinfo{year}{2021}), \bibinfo{pages}{216}.
\newblock
\showISSN{2047-7538}
\urldef\tempurl%
\url{https://doi.org/10.1038/s41377-021-00658-8}
\showDOI{\tempurl}


\bibitem[Yaqoob et~al\mbox{.}(2020)]%
        {360ABRSurvey}
\bibfield{author}{\bibinfo{person}{Abid Yaqoob}, \bibinfo{person}{Ting Bi}, {and} \bibinfo{person}{Gabriel-Miro Muntean}.} \bibinfo{year}{2020}\natexlab{}.
\newblock \showarticletitle{A Survey on Adaptive 360° Video Streaming: Solutions, Challenges and Opportunities}.
\newblock \bibinfo{journal}{\emph{IEEE Communications Surveys \& Tutorials}} \bibinfo{volume}{22}, \bibinfo{number}{4} (\bibinfo{year}{2020}), \bibinfo{pages}{2801--2838}.
\newblock
\urldef\tempurl%
\url{https://doi.org/10.1109/COMST.2020.3006999}
\showDOI{\tempurl}


\bibitem[Zaki et~al\mbox{.}(2015)]%
        {queuingdelaygradient3}
\bibfield{author}{\bibinfo{person}{Yasir Zaki}, \bibinfo{person}{Thomas P\"{o}tsch}, \bibinfo{person}{Jay Chen}, \bibinfo{person}{Lakshminarayanan Subramanian}, {and} \bibinfo{person}{Carmelita G\"{o}rg}.} \bibinfo{year}{2015}\natexlab{}.
\newblock \showarticletitle{Adaptive Congestion Control for Unpredictable Cellular Networks}.
\newblock \bibinfo{journal}{\emph{SIGCOMM Comput. Commun. Rev.}} \bibinfo{volume}{45}, \bibinfo{number}{4} (\bibinfo{date}{Aug.} \bibinfo{year}{2015}), \bibinfo{pages}{509–522}.
\newblock
\showISSN{0146-4833}
\urldef\tempurl%
\url{https://doi.org/10.1145/2829988.2787498}
\showDOI{\tempurl}


\bibitem[Zhu and Pan(2013)]%
        {NADA}
\bibfield{author}{\bibinfo{person}{Xiaoqing Zhu} {and} \bibinfo{person}{Rong Pan}.} \bibinfo{year}{2013}\natexlab{}.
\newblock \showarticletitle{NADA: A Unified Congestion Control Scheme for Low-Latency Interactive Video}. In \bibinfo{booktitle}{\emph{2013 20th International Packet Video Workshop}}. \bibinfo{pages}{1--8}.
\newblock
\urldef\tempurl%
\url{https://doi.org/10.1109/PV.2013.6691448}
\showDOI{\tempurl}


\end{thebibliography}

% \appendix
\appendix

\section{Illustration of Foveated Spatial Compression/Decompression}
\label{app:sec1}

\begin{figure}[h]
    \centering
    \begin{subfigure}[b]{.3\textwidth}
        \includegraphics[width=1.03\linewidth]{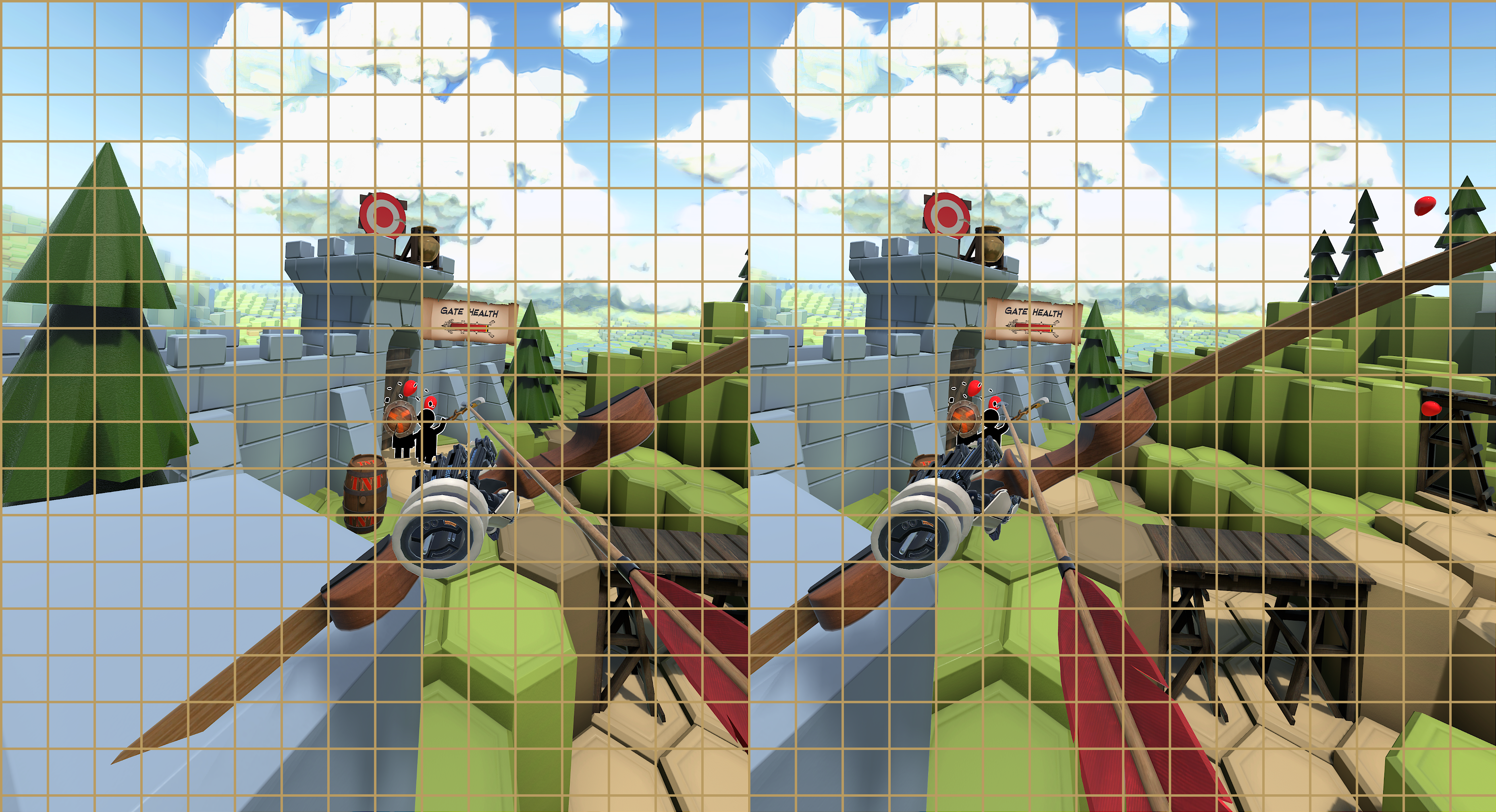}
        \caption{Game frame (3712X2016)}
    \label{fig:oframe_vis}
    \end{subfigure}
    \hfill
     \begin{subfigure}[b]{.3\textwidth}
        \includegraphics[width=1.03\linewidth]{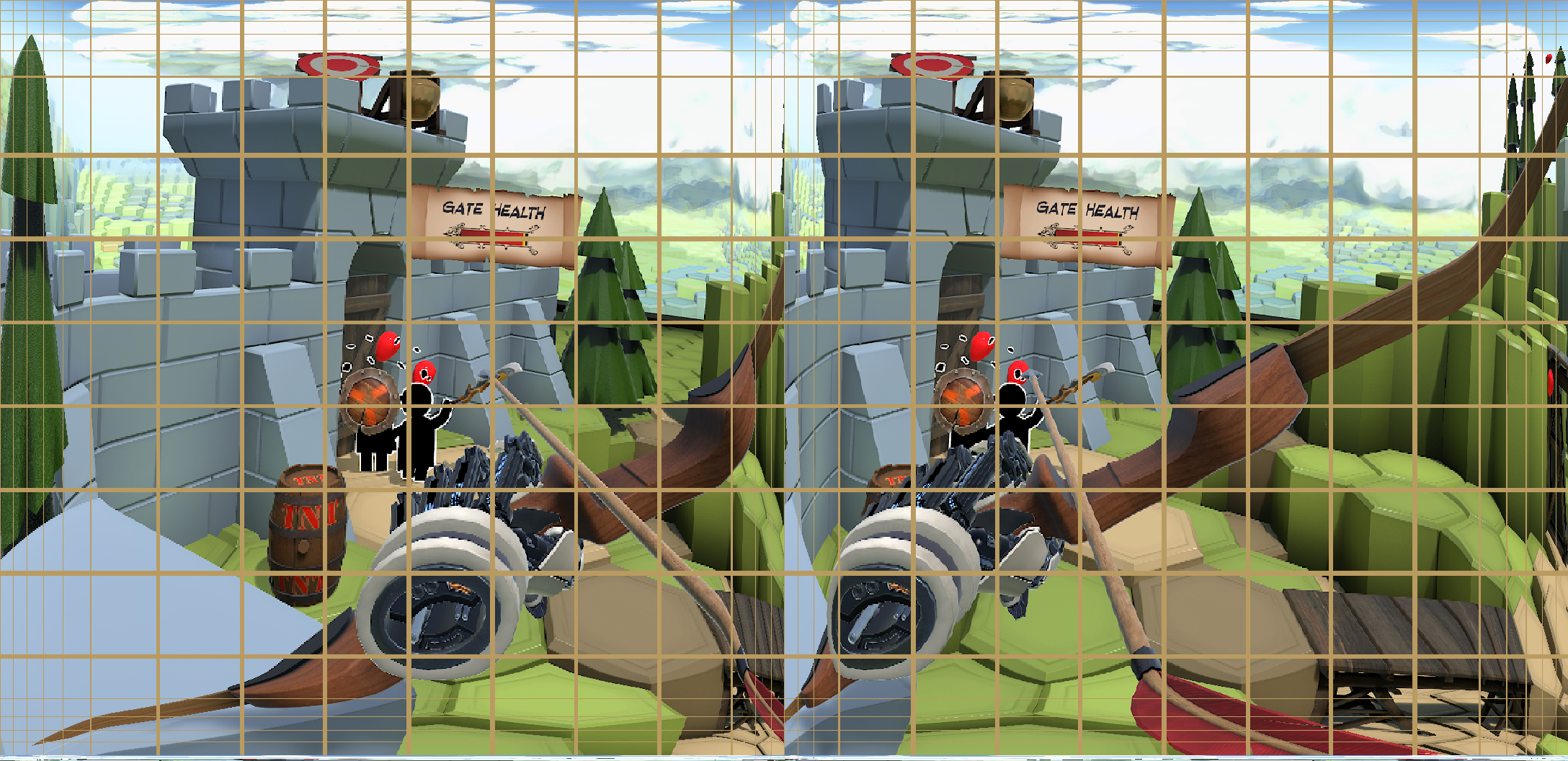}
        \caption{FSC frame (2176x1056)}
    \label{fig:rframe_vis}
    \end{subfigure}
    \hfill
     \begin{subfigure}[b]{.3\textwidth}
        \includegraphics[width=1.03\linewidth]{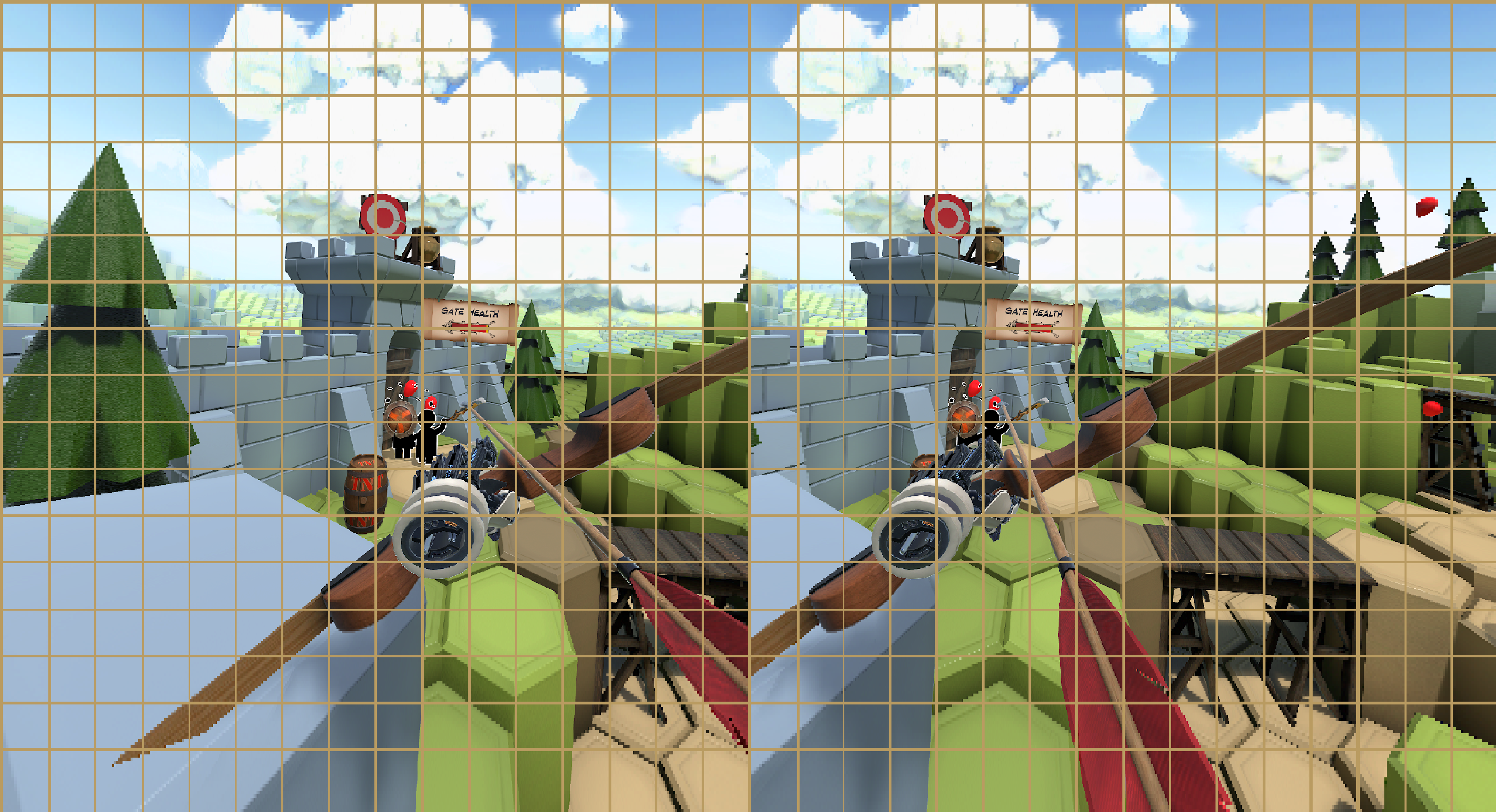}
        \caption{FSD frame (3712x2016)}
    \label{fig:backframe_vis}
    \end{subfigure}
    \caption{Visual representation of the frame produced by foveated spatial compression (FSC) and dynamic foveated spatial decompression (FSD).}
    \label{fig:fsc_process_vis}
\end{figure}

\autoref{fig:fsc_process_vis} illustrates the process of foveated spatial compression (FSC) and foveated spatial decompression (FSD).  \autoref{fig:oframe_vis} depicts the original game frame that is rendered by the game engine on the server. The game frame serves as input to FSC as a post-rendering process. FSC reduces resolution by spatially compressing pixels in the peripheral region while preserving pixel density in the fovea region. This results in an FSC frame, shown in \autoref{fig:rframe_vis}, with a grid overlay to visualize the compression effect. It demonstrates that the compression ratio increases with distance from the center. The FSC frame is then encoded and transmitted to the client. Upon receiving and decoding the FSC frame, the client restores the game frame at the original resolution using FSD function, as shown in \autoref{fig:backframe_vis}. The fovea region maintains high quality, while the peripheral region experiences a slight, acceptable drop in quality due to the lossy nature of FSC. This reduction is not noticeable, as it is limited to the peripheral area.

% \begin{figure}[H]
%     \centering
%         \centering
%         \includegraphics[width=.95\linewidth]{figures/FR_Frame_with_grid.png} % Path to diagram c
%         \caption{Non-uniform spatial compression and generated FSC frame}
%         \label{fig:FSC_frame}
% \end{figure}

\section{Foveation Controller and Quantization Curve}
\label{app:sec2}
\begin{figure}[H]
    \centering
    \includegraphics[width=.6\linewidth]{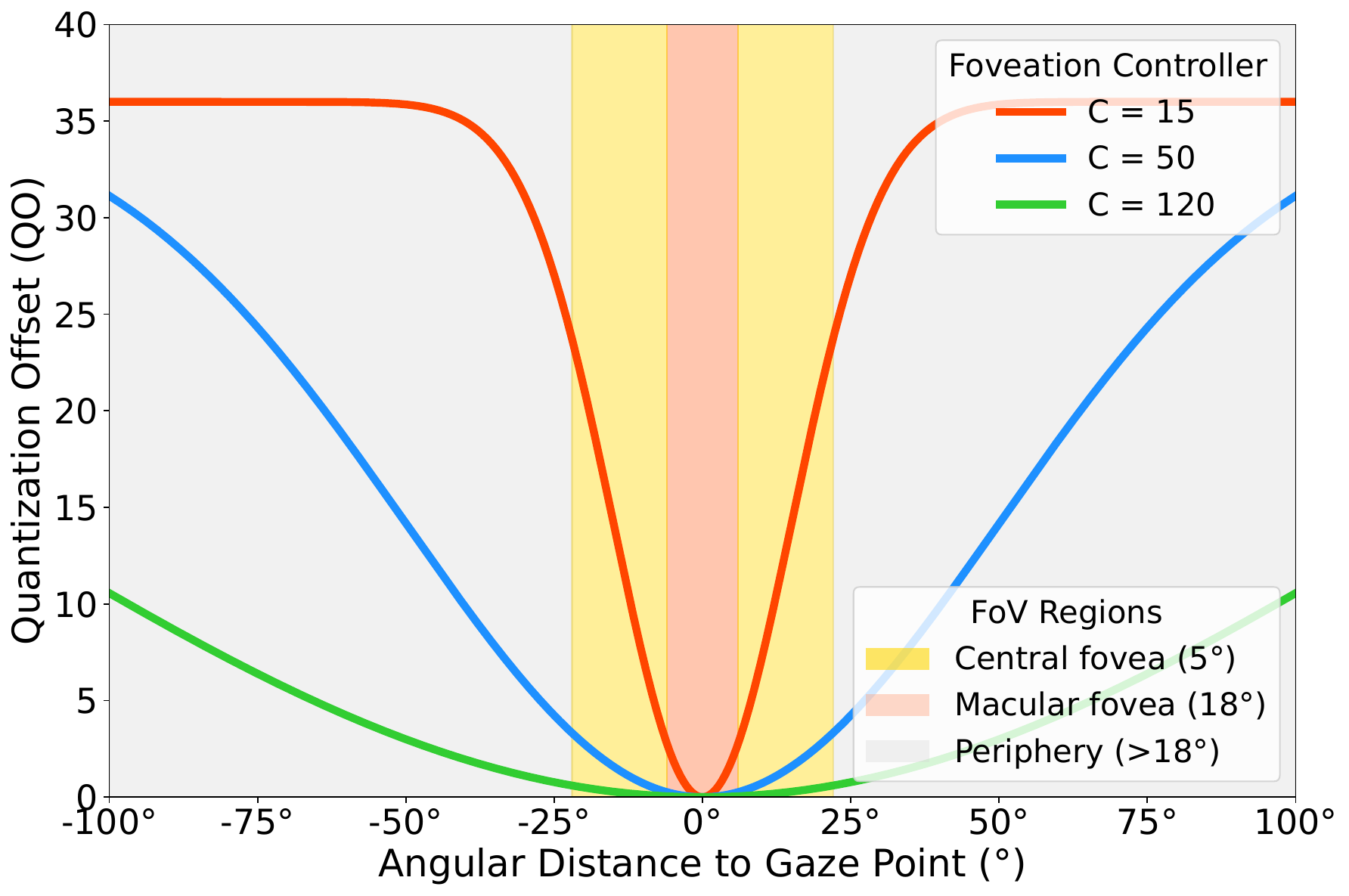}
    \caption{Quantization Offset (QO) across FoV regions and various values of foveation controller C, with central fovea span 5\degree, macular 18\degree, and periphery >18\degree \cite{degreesforfovea}.}
    \label{fig:foveation_controller}
\end{figure}
\sysname computes QO and assigns QP values non-uniformly to each macroblock using the foveation controller \(C\), which adjusts to expand or shrink the foveation region, as shown in \autoref{fig:foveation_controller}. Specifically, as  \(C\) increases, the standard deviation of the Gaussian curve increases, leading to a larger foveation region with a lower QP, while a decrease in  \(C\) narrows the curve, resulting in a smaller foveation region with a lower QP. \sysname utilizes a finite state machine to control \(C\), adjusting the transmission bitrate in response to fluctuations in network bandwidth. The C value ranges between predefined $C_{min}$ and $C_{max}$. Notably, the configuration of \(C_{min}\) and \(C_{max}\) depends on the frame resolution. 

\section{Ablation Study:Latency Comparison Across Devices for High Resolutions}
\label{app:sec3}
% \subsection{Latency Comparison Across Devices for High Resolutions}

\begin{figure}[H]
    \centering
    \begin{subfigure}[b]{.85\textwidth}
        \includegraphics[width=\linewidth]{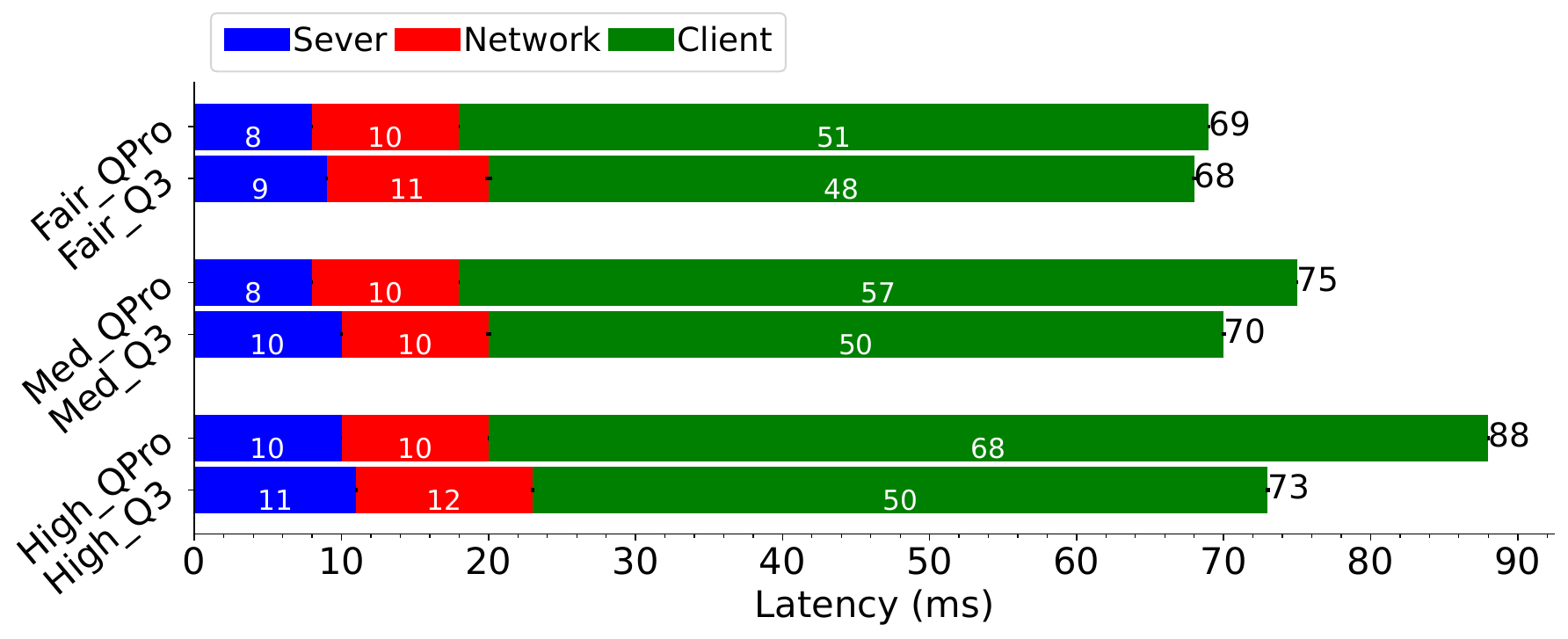}
        \caption{MTP latency across its sources}
        \label{fig:adx_high_res_devices_sources}
    \end{subfigure}
    \begin{subfigure}[b]{.85\textwidth}
        \includegraphics[width=\linewidth]{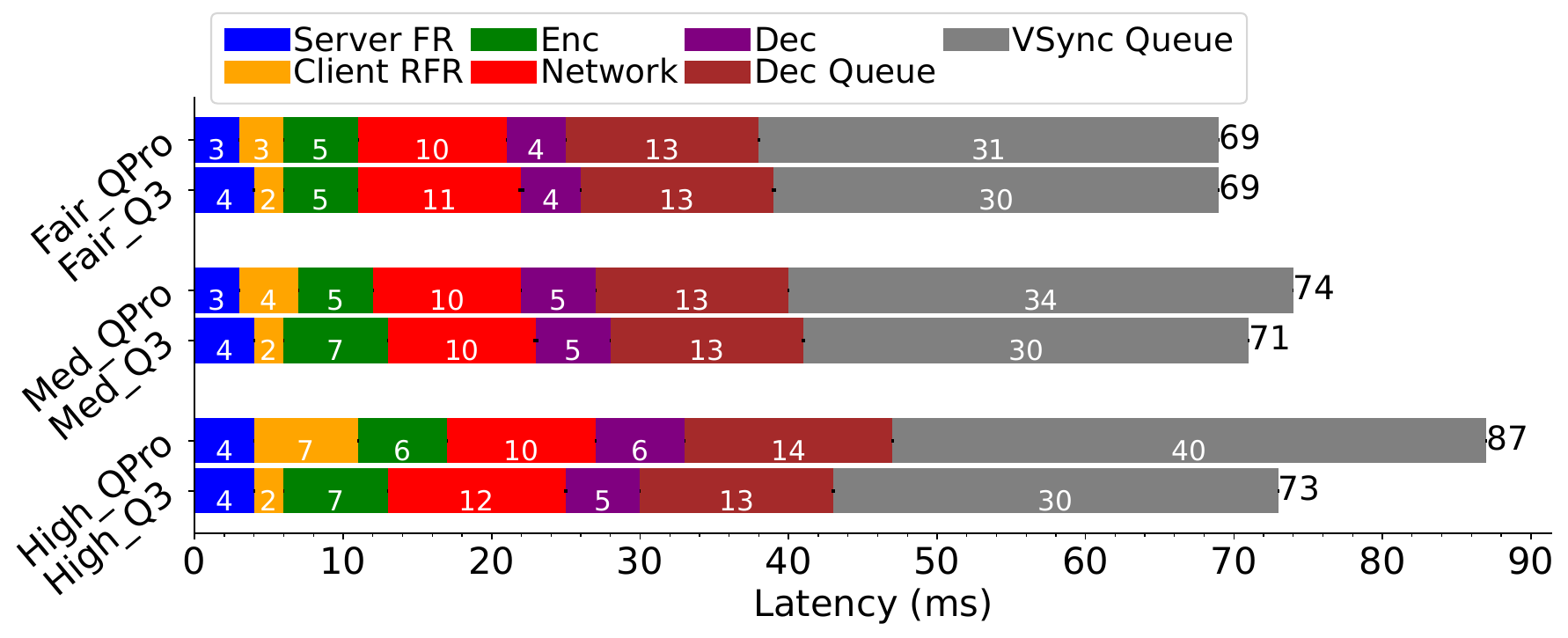}
        \caption{Breakdown of MTP latency across the pipeline}
        \label{fig:adx_high_res_devices_breakdown_latency_pipeline}
    \end{subfigure}
    \caption{Breakdown of MTP latency for higher resolutions using Quest Pro and Quest 3, with resolutions [High \((5184\times2848)\), Medium \((4288\times2336)\), Fair \((3712\times2016)\)] }
     \label{fig:adx_high_res_devices}
\end{figure}

% \begin{figure}[H]
%     \centering
    
%     \includegraphics[width=\linewidth]{figures/adx_high_res_devices.pdf}
%     \caption{Breakdown of MTP latency for higher resolutions using Quest Pro and Quest 3, with resolutions [High \((5184\times2848)\), Medium \((4288\times2336)\), Fair \((3712\times2016)\)]}
%     \label{fig:adx_high_res_devices}
% \end{figure}
% \begin{figure}[H]
%     \centering
%     \includegraphics[width=\linewidth]{figures/adx_high_res_devices_breakdown_latency.pdf}
%     \caption{Breakdown of MTP latency for higher resolutions using Quest 3 and Quest Pro, with resolutions [High \((5184\times2848)\), Medium \((4288\times2336)\), Fair \((3712\times2016)\)]}
%     \label{fig:adx_high_res_devices_breakdown_latency}
% \end{figure}
We conduct further experiments with two VR headsets to assess additional client latency at higher resolutions, comparing the performance of Quest Pro and Meta Quest 3. We used Quest Pro in the previous evaluation because it supports built-in gaze tracking. Since Quest 3 lacks this feature, we use eye gaze data collected during the experiments on Quest Pro. We replicate the same experiment for \sysname on Quest 3 by processing the real-world eye gaze data points. As shown in \autoref{fig:adx_high_res_devices_breakdown_latency_pipeline}, \sysname reduces client-side latency by 3 ms for Fair resolution, 7 ms for Medium, and 18 ms for High resolution when using Quest 3. The results confirm our interpretation in Section 6.4, attributing the additional increase in MTP latency to the resources of the Quest Pro. Specifically, with the advanced hardware of Quest 3, \sysname achieves end-to-end latency mean values of 68 ms for Fair resolution, 70 ms for Medium, and 73 ms for High resolution. Although MTP latency increases with higher resolutions, the increment is negligible and often imperceptible to users. As a result, \sysname maintains low end-to-end latency regardless of resolution and client hardware specifications.

\section{Bitrate Adaptability} 
\label{app:bitrate adaptivity}
We evaluate adaptability to available bandwidth using the bitrate imbalance metric, which measures the difference between the server-side sending bitrate and the client-side receiving bitrate. This metric indicates whether there is overshooting (sending more than the client can receive) or undershooting (sending less than the client can receive). \sysname achieves the lowest bitrate difference ($0.76\pm0.08$) in game streaming over mobile network setup, followed by FovOptix ($2.46\pm0.5$) and GCC ($4.1\pm1$). ALVR and CGFVE present the highest bitrate difference, averaging $31.5\pm1.2$ and $54\pm26$, respectively. In WiFi setup, the benchmarks demonstrate comparable performance, with \sysname achieving the lowest ($0.3\pm0.03$), followed by FovOptix ($1.01\pm0.1$) and GCC ($1.013\pm1$), while ALVR and CGFVE present a slightly higher bitrate difference, $1.06\pm0.26$ and $1.46\pm0.39$, respectively. 

\section{Algorithm of Foveated Spatial Decompression (FSD) Function}
\label{app:sec4}
The pseudocode for spatial decompression along the x-axis is shown in Algorithm~\autoref{algo:DFSD}, with a similar process applicable to the y-axis. This function is primarily used in Gaze Mapping (\autoref{sec:video_encoding}) and client-side rendering (\autoref{sec:fvd_rr}). The FSD algorithm acts as an inverse function of FSC mentioned in Algorithm~\autoref{algo:DFSC}, allowing us to expand the compressed image from compressed frame $(W_r, H_r)$ to the dimensions of the game frame $(W_o, H_o)$.
\begin{algorithm}[h]
    \caption{Dynamic Foveated Spatial Decompression (FSD)}
    \begin{algorithmic}[1]
        \State\textbf{Input:} $\text{FSC frame}, W_r, W_o, X_o, X_{size}, X_{comp}$
        \State\textbf{Output:} Game frame 
        \State $bound_{left} \gets 2(1-X_{size})*\frac{X_o}{W_o}$
        \State $bound_{right} \gets (1-X_{size})*\frac{X_o-W_o}{W_o}+1$
        \State \text{Initialize Reconstructed Game frame}
        \For{$i$ from $0$ to $W_o$}
        \If{$i<bound_{left}$}\Comment{Peripheral region}
        \State $i' \gets \frac{i}{W_o}*\frac{W_r}{((X_{comp}-1)*X_{size}+1)}$ 
        \State $i' \gets \frac{i}{X_{comp}}$
        \ElsIf{$i>bound_{right}$}\Comment{Peripheral region}
        \State $i' \gets \frac{i-W_o}{X_{comp}} +W_r$
        \Else\Comment{Fovea region}
        \State $c_f \gets \frac{X_{comp}-1}{X_{comp}}*(1-X_{size})*(\frac{X_o}{W_o})$
        \State $i' \gets i - c_f*W_o$
        \EndIf
        \State \text{Reconstructed game frame}$[i] \gets \text{FSC frame}[i']$
        \EndFor
        % \State\Return \text{Reconstructed game frame}
    \end{algorithmic}
    \label{algo:DFSD}
\end{algorithm}

\section{Algorithm of Gaze-driven Encoding}
\label{app:sec5}
The quality manager generates a QP map using the foveation controller $C$ and gaze location $(X_o, Y_o)$. The QP map contains QP values that are lower around the gaze location and increase gradually, following the bell curve. The $C$ is dynamically adjusted to extend or shrink the foveation region in response to the bandwidth usage. The process flow of gaze-driven FVE is presented as pseudocode in Algorithm~\autoref{algo:gaze_drivernFVE}:
\begin{algorithm}
    \caption{Dynamic Foveated Encoding}
    \begin{algorithmic}[1]
        \State \textbf{Input: } $C, (W_r, H_r), (X_o, Y_o), [QO_{min},QO_{max}]$
        \State \textbf{Output: } QP Map
        \State $X_r, Y_r \gets \text{FSD}(X_o, Y_o)$ \
        \State $W_{QP}, H_{QP} \gets \left\lceil \frac{W_r}{16} \right\rceil, \left\lceil \frac{H_r}{16} \right\rceil$ \Comment{QP Map Width and Height}
        \State $X_{QP}, Y_{QP} \gets \left\lceil \frac{X_r}{16} \right\rceil, \left\lceil \frac{Y_r}{16} \right\rceil$ 
        \State Initialize QP Map with $(X_{QP}, Y_{QP})$ 
        \For{every $(i, j)$ in QP map}
            \State QO $\gets QO_{max}-QO_{max}\times\exp{-\frac{\sqrt{(i-X_{QP})^2+(j-Y_{QP})^2}}{2C^2}}$ 
            \State QP$[i,j] \gets QP_{const}+QO$
        \EndFor
    \end{algorithmic}
    \label{algo:gaze_drivernFVE}
\end{algorithm}

\end{document}